\documentclass{iopart}
\pdfoutput=1
\usepackage{graphicx}
\usepackage{float}
\usepackage{amssymb}

\newcommand{\ket}[1]{| #1 \rangle}

\begin{document}

\title{Loschmidt  echo and Poincar\'e recurrences of entanglement}

\author{Leonardo Ermann$^{1,2}$, Klaus M.Frahm$^3$,\\  
        and Dima L.Shepelyansky$^3$}
\address{$^1$Departamento de F\'{\i}sica Te\'orica, GIyA,
         Comisi\'on Nacional de Energ\'{\i}a At\'omica, 
         Av.~del Libertador 8250,   1429 Buenos Aires, Argentina\\
$^2$Consejo Nacional de Investigaciones
         Cient\'ificas y T\'ecnicas (CONICET), C1425FQB, Buenos Aires, Argentina\\
$^3$Laboratoire de Physique Th\'eorique du CNRS, 
Universit\'e de Toulouse, UPS, 31062 Toulouse, France}
 \date{January ??, 2022}
 

\begin{abstract}
We study numerically the properties of entanglement of two
interacting, or noninteracting, particles evolving in
a regime of quantum chaos in the quantum Chirikov standard map.
Such pairs can be viewed as interacting, on noninteracting,
Einstein-Podolsky-Rosen pairs
in a regime of quantum chaos.
The analysis is done with such tools as  the Loschmidt echo of entanglement
and  the Poincar\'e recurrences of entanglement in presence of absorption.
The obtained results show unusual features of the entropy of entanglement
and the spectrum of Schmidt decomposition
with their dependence on interactions at different quantum chaos regimes.
\end{abstract}

\maketitle
\section{Introduction}
The ancient dispute between Loschmidt and Boltzmann about emergence of statistical laws
from time reversible dynamical equations \cite{bolzmann1,loschmidt,bolzmann2} 
(see also \cite{mayer})
found its modern resolution on the basis of
phenomenon of dynamical chaos with its exponential instability of trajectories
which breaks the time reversal in presence of exponentially
small errors \cite{arnold,sinai,chirikov1979,lichtenberg}.
In quantum mechanics this exponential instability of chaos has been shown to exist
only on a logarithmically short 
Ehrenfest time scale \cite{chi1981,dls1981,chi1988,ehrenfestime}:
\begin{equation}
\label{eqehrenfest}
t_E \sim (\ln q)/h = \ln(I/\hbar) /h  
\end{equation}
where $q = I/\hbar $ is a typical quantum number, $I$ is a corresponding classical action,
$\hbar$ is the Planck constant and $h \geq \Lambda$ is the Kolmogorov-Sinai entropy 
\cite{arnold,sinai,chirikov1979,lichtenberg}
which is larger or equal to the Lyapunov exponent of a dynamical chaotic trajectory.
This time $t_E$ is so short due to an exponentially rapid spreading of minimal
coherent wave packet so that after this time the Ehrenfest theorem \cite{ehrenfest1927}
looses its validity. 

Various properties of quantum chaos of one-particle 
quantum evolution, which is chaotic in the classical limit, are described and reviewed in 
\cite{chi1981,chi1988,gutzwiller,haake}. While the classical chaotic dynamics breaks time reversal
due to the exponential growth of errors, in \cite{dls1983} it was shown that 
in the regime of quantum chaos the time reversal remains stable even if the 
numerical simulations are done on the same computer for classical and quantum
evolution. This result was obtained for  the Chirikov standard map
which describes the generic features of chaotic dynamics with divided phase space
\cite{chirikov1979,lichtenberg,stmap}. The studies of effects of Hamiltonian perturbations 
acting on the quantum evolution during the return path of 
time reversal have been extended in \cite{peres}
and the decoherence effects for this Loschmidt echo  have been analyzed in \cite{jalabert}
with links to the Lyapunov exponent.
Various interesting properties of Loschmidt echo have been studied by different groups
being described in \cite{prosen,jacquod,jalabertsch,pastawski}.
In the context of quantum computing the properties of fidelity and Loschmidt echo
for time reversal were reported in \cite{georgeot1,georgeot2}.
The time reversal of atomic Bose-Einstein condensate in the regime of quantum chaos of the
Chirikov standard map was experimentally realized by the Hoogerland group \cite{hoogerland},
following the theoretical proposal \cite{martin}.

The above studies of the time reversal and Loschmidt echo are done for
one particle quantum  evolution. However, it is also interesting to
analyze the properties of entanglement in systems of quantum chaos.
Indeed, the  fundamental work of Einstein-Podolsky-Rosen (EPR) \cite{epr}
about a distant entanglement \cite{schrodinger}  of a pair of noninteracting
distinguishable particles is now at the foundations of
quantum information and communications \cite{chuang,karol,deutsch}.

Recently, the properties of chaotic EPR pairs without interactions,
the effects of time reversal and measurements were analyzed 
for the quantum Chirikov map in \cite{eprchaos}. 
Here, the Schmidt decomposition  of the EPR wavefunction 
\cite{schmidt} (see also \cite{fedorov})
is found to be especially useful. Without interactions the entropy of entanglement $S$
of the EPR pair \cite{chuang,karol} is preserved during a quantum evolution.
Thus it is interesting to  study how this quantity 
and Loschmidt echo $M(t)$ are affected by interactions between
particles. With this aim we present here the analysis of these
quantities for chaotic EPR pairs with interactions
in the quantum Chirikov standard map. 
This model was already investigated in \cite{dlstip,borgonovitip} 
in the context of interaction effects on the dynamical localization 
but the entropy of entanglement was not studied there.

In addition, we also study how the entropy of entanglement $S$ 
in this model is affected by the absorption of one or two particles.
In a certain sense the absorption can be considered as
some kind of measurement and it is interesting to understand
its influence on EPR characteristics.
We note that in the case of quantum evolution of one particle
the effects of absorption have been studied in this system
in \cite{borgonovi1,maspero1,sokolovabsorb,frahmabsorb,maspero2,dlsweyl}.
The probability that a particle remains inside 
the system can be considered as a quantum version of Poincar\'e
recurrences \cite{poincare} which in the classical case of fully chaotic system 
decays exponentially with time while in the case of divided phase space
with stability islands  the decay is algebraic 
(see \cite{chsh,frahmulam} and Refs. therein).
Thus, in this work we study the decay of Poincar\'e recurrences 
of entanglement for a chaotic EPR pair with and without interactions. 

This paper is constructed as follows: the model is described in Section 2,
the Loschmidt echo of entanglement is studied in Section 3,
various cases of Poincar\'e recurrences of entanglement are analyzed in
Sections 4, 5, 6 and discussion of the results is given in Section 7
(specific points are presented in the Appendix
and additional data are given in Supplementary Material (SupMat)).

\section{Model description}
For one particle the classical dynamics is described 
by the Chirikov standard map \cite{chirikov1979}:
\begin{equation}
\label{stmap}
\bar{p} = p + k \sin{ x} \; , \;\; 
\bar{x} = x + T \bar{p} \; .
\end{equation}
Here $x$ represents a cyclic variable $0 \leq x < 2\pi$
for the case of the kicked rotator,
 $p$ is the particle momentum.
The bars denote the new values of variables 
after one iteration of this  symplectic map. 
The  dynamics depends on a single chaos parameter
$K=kT$ with a transition from integrability to unlimited chaotic
diffusion in momentum for $K > K_c =0.9715...$ \cite{chirikov1979,lichtenberg}.
The system dynamics is reversible in time, e.g. by
inverting all velocities in a middle of free rotation between two kicks \cite{dls1983}.
This map captures the generic properties of chaos in systems with
integrable islands surrounded by chaotic components, its applications
to various physical systems are summarized in \cite{stmap}.

The dynamics in a chaotic component has a positive Kolmogorov-Sinai entropy
$h$ which characterizes the exponential divergence of nearby trajectories.
For $K > 4$ we have  $h \approx \ln(K/2)$ \cite{chirikov1979}.
For $K > K_c$ there is an unlimited diffusive momentum growth with time $t$, measured
in number of map iterations:
$\langle(\Delta p)^2\rangle = 2 D t$ with a diffusion coefficient
$D \approx k^2/4$ (see more details in  \cite{chirikov1979,dls1987}).

The quantum  state $\ket{\psi}$  propagation over a period is given 
by a unitary operator (here still for one particle) 
 $U_{\rm KR}^{(1)}$ \cite{chi1981,chi1988}:
\begin{eqnarray} 
\label{qmap}
\ket{\bar{\psi}} = U_{\rm KR}^{(1)} \ket{\psi} = 
e^{-iT\hat{p}^2/2} e^{-ik\cos{\hat{x}}} \ket{\psi} \; .
\end{eqnarray} 
Here the momentum $p$ is measured in recoil units of the optical lattice with 
$\hat{p}=-i \partial / \partial x $. Hence, $T=\hbar$ plays the role of
an effective dimensionless Planck constant and the classical limit
corresponds to $T=\hbar \rightarrow 0$, $k \rightarrow \infty$,
$K=kT = const$. Here we consider  the case of a kicked rotator
with a wave function (in position representation) 
$\psi(x)=\langle x\ket{\psi}$ being 
periodic on a circle $\psi(x+2\pi)=\psi(x)$.
In this case the free rotation corresponds (in momentum representation) 
to the phase shift
${\bar {\psi}}_n = \exp(-iTn^2/2) \psi_n$ with 
$\psi_n=\langle p\ket{\psi}$ being the wave function 
(in momentum representation) at $p=n$. 
The effects of quantum interference lead to dynamical localization
of chaotic diffusion on a time scale $t_D \approx D/\hbar^2 \gg t_E$
and an exponential localization of quasienergy eigenstates
with a localization length $\ell = D/(\hbar^2) \approx k^2/4$ 
\cite{dls1987,chi1988,fishman1,fishman2}.
This dynamical localization is similar to the Anderson localization
of electrons in a disordered solid \cite{anderson}
and it has been observed in experiments with cold atoms in 
kicked optical lattices in \cite{raizen,garreau}. The time reversal of 
atomic waves in this system has been realized in \cite{hoogerland}
following the theoretical proposal \cite{martin}.

For two noninteracting or interacting particles
the evolution  operator $U_{\rm KR}$ is given by:
\begin{eqnarray}
\label{eqUKR1}
U_{\rm KR}&=&\exp(-iT(\hat p_1^2+\hat p_2^2)/4+i\hat U/2)\times\\
\label{eqUKR2}
&&\exp(-ik\cos\hat\theta_1-ik\cos\hat\theta_2)\times\\
\label{eqUKR3}
&&\exp(-iT(\hat p_1^2+\hat p_2^2)/4+i\hat U/2)\ .
\end{eqnarray}
Here $\hat U$ is the interaction operator which is diagonal 
in momentum representation with eigenvalues being either 
$U$ for $p_1=p_2$, $U/2$ for $0<|p_1-p_2|\le U_r$ or $0$ for other cases 
with $|p_1-p_2|> U_r$ with $p_{1,2}=n_{1,2}$ being momentum of first and second particle. 
Here $U$ is the interaction strength parameter 
and $U_r$ is the interaction range 
(chosen as $U_r=0$ or $U_r=1$ in this work).
At $U_r=0$ we have the case of Hubbard on-site interaction
which was first studied in \cite{dlstip,borgonovitip} for the kicked rotator. 
The difference $|p_1-p_2|$ is computed with respect to the periodic boundary 
conditions, i.e., $|p_1-p_2|=1$ if for example $p_1=N/2-1$ and $p_2=-N/2$
and similarly for other cases; 
$\hat\theta_j$ represents the usual phase operator of particle $j$. 

We also mention that in (\ref{eqUKR1}-\ref{eqUKR3}), we use, in 
contrast to (\ref{qmap}), the symmetrized version of the map where a half 
free rotation is applied before and after the kick operator. This point 
is important to keep the time reversal symmetry and in particular for the 
studies in the next section.

\section{Loschmidt echo of entanglement}

In this section we consider the case of Hubbard on-site 
interaction of the quantum kicked rotator given by 
Eqs.~(\ref{eqUKR1}),(\ref{eqUKR2}),(\ref{eqUKR3}) with $U_r=0$ and $U=2$.
The other system parameter values are $N=2^{10}=1024$, 
$T=\hbar=\epsilon=5/8$, and $K=5$ such that $k=K/\hbar=8$. 
As initial state we take a non-entangled state with $p_1=0$, $p_2=1$:
\begin{equation}
\ket{\psi(t=0)}=\ket{p_1=0} \otimes \ket{p_2=1}\ .
\end{equation}

The entropy of entanglement is given by (see e.g. \cite{chuang}; 
see also the next section and Appendix A.1 for additional details
on the entropy computation and the numerical method for the quantum 
time evolution)
\begin{equation}
\label{eqentropydef1}
S(t)=-\mbox{Tr}[\rho_1\log_2(\rho_1)]
\end{equation}
where $\rho_1$ is the reduced density matrix for the first particle 
obtained by a trace over the second particle.

In order to study the Loschmidt echo, we compute the time evolution 
forward in time with parameter values $U_f$ and $k_f$: 
$U_{\rm KR}(U_f,k_f)$ till $t=t_r$ where the time reversal takes place
and then backward in time $U_{\rm KR}(U_b,k_b)$ till reaching $t=2 t_r$.
We analyze the cases of $U_f=U$, 
$U_b=U_f+\Delta U$, $k_b=k_f=k$; and $U_f=U_b=U$, 
$k_f=k$, $k_b=k_f+\Delta k$. 

\begin{figure}[h]
\begin{center}
\includegraphics[width=0.8\textwidth]{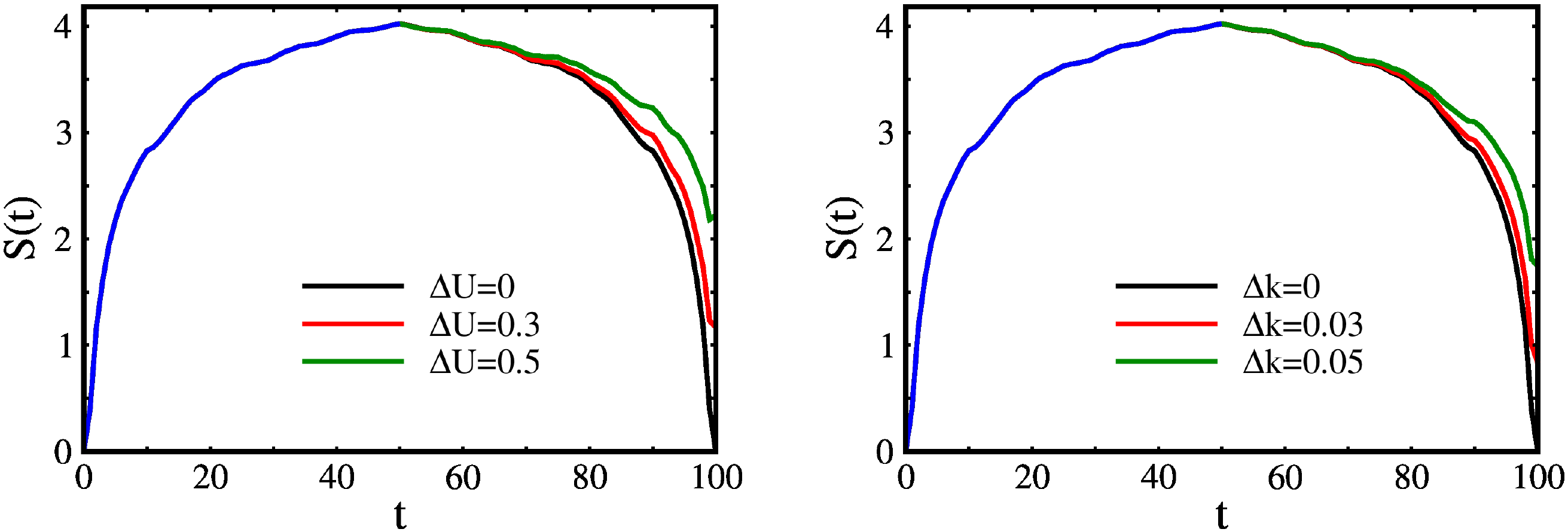}
\caption{Time dependence of the entropy of entanglement $S$ for the 
initial state $\vert \psi(t=0) \rangle=\vert p_1=0\rangle \otimes \vert p_2=1\rangle$ 
with time evolution given by the quantum 
Chirikov standard map. The time reversal is performed 
after $t_r=50$ quantum map iterations.
The blue curve in both panels show the forward time 
evolution  $0\geq t \geq t_r$, the black curves show 
the backward time evolution $t_r\geq t\geq 2t_r=100$ 
with the exact time reversal using $T=4\pi-\epsilon$.
{\em Left panel:} Curves of other colors show the 
backward time evolution with perturbation $U_b=U_f+\Delta U$ 
at $\Delta U=0.3$ (red curve) and $\Delta U=0.5$  
(blue curve).
{\em Right right:} Curves of other colors show the 
backward time evolution with $k_b=k_f+\Delta k$ 
at $\Delta k=0.03$ (red curve) and $\Delta k=0.05$ (blue curve);
$U_f=U_b=U$.
The system parameters are: $N=1024$, 
$T=\hbar=\epsilon=5/8$, $U=U_f=2$, and $k=K/\hbar=8$.
}
\label{fig1}
\end{center}
\end{figure}

Fig.~\ref{fig1} shows the time dependence of the entropy of entanglement
(\ref{eqentropydef1}) with $t_r=50$ and:
 $U_f=U=2$, $U_b=U+\Delta U$ and $k_b=k_f=k=8$ with 
 $\Delta U=0,0.3,0.5$ in left panel; and $U_f=U_b=2$, $k_f=k=8$ and 
 $k_b=k+\Delta k$ with $\Delta k=0,0.03,0.05$.
We have verified that a further increase of $N$
to values of $10^{11}=2048$ and $10^{12}=4096$ provide 
identical results up to numerical round-off errors.
The results show that finite perturbations
$\Delta U$ or $\Delta k$ break time reversal of
entropy of entanglement $S(2t_r)$. 

\begin{figure}[h]
\begin{center}
\includegraphics[width=0.8\textwidth]{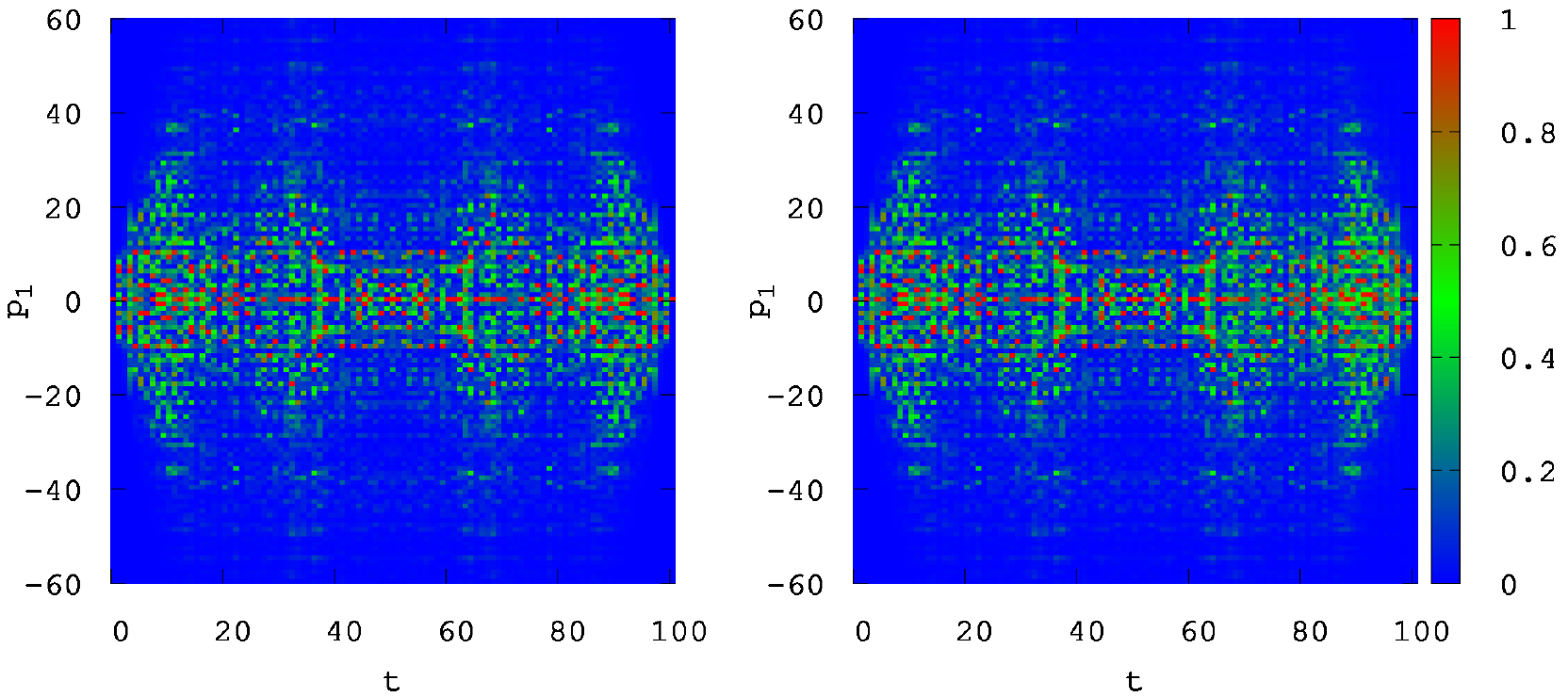}
\caption{Time evolution of probability of
the first particle $w(p_1,t)$ (color density plot) 
for $U=0$, $t_r=50$, $N=1024$, 
$\hbar_{eff}=\epsilon=5/8$, $k=K_{eff}/\hbar_{eff}=8$ and $-60\leq p_1\leq 60$ (y-axis), 
$0\leq t \leq 100$ 
(x-axis). 
The color bar gives values of
$[w(p_1,t)/w_{\rm max}(t)]$ with 
$w_{\rm max}(t)=\max_{p_1}w(p_1,t)$ 
being the density maximum at a given value of $t$.
{\em Left panel} shows the case of $\Delta U=0$  
(corresponding to the black curve of Fig.\ref{fig1} left panel);
{\em right panel} represents the case of $\Delta U=0.5$ 
(corresponding to the green curve of Fig.\ref{fig1} left panel). 
}
\label{fig2}
\end{center}
\end{figure}

The probability distributions in momentum of the first particle ($p_1=n_1$)
$w(p_1,t)=\langle p_1\vert\rho_1(t)\vert p_1\rangle$,
taken at different moments in time, are presented
in color in Fig.~\ref{fig2} for $\Delta U=0$ (left panel) 
and $\Delta U=0.5$ (right panel) with $t_r=50$ and the same other 
parameter values of Fig.~\ref{fig1}. 
The color bar scale correspond
to $[w(p_1,t)/w_{\rm max}(t)]$ with 
$w_{\rm max}(t)=\max_{p_1}w(p_1,t)$ 
being the density maximum at a given value of $t$.
We see that the perturbation $\Delta U$
is relatively weak and the global profile of density
distribution $w(p_1,t)$ is only weakly perturbed
as compared to the case of exact time reversal.
However, in the next figures we show that
the echo characteristics are more sensitive
to perturbations.

\begin{figure}[h]
\begin{center}
\includegraphics[width=0.8\textwidth]{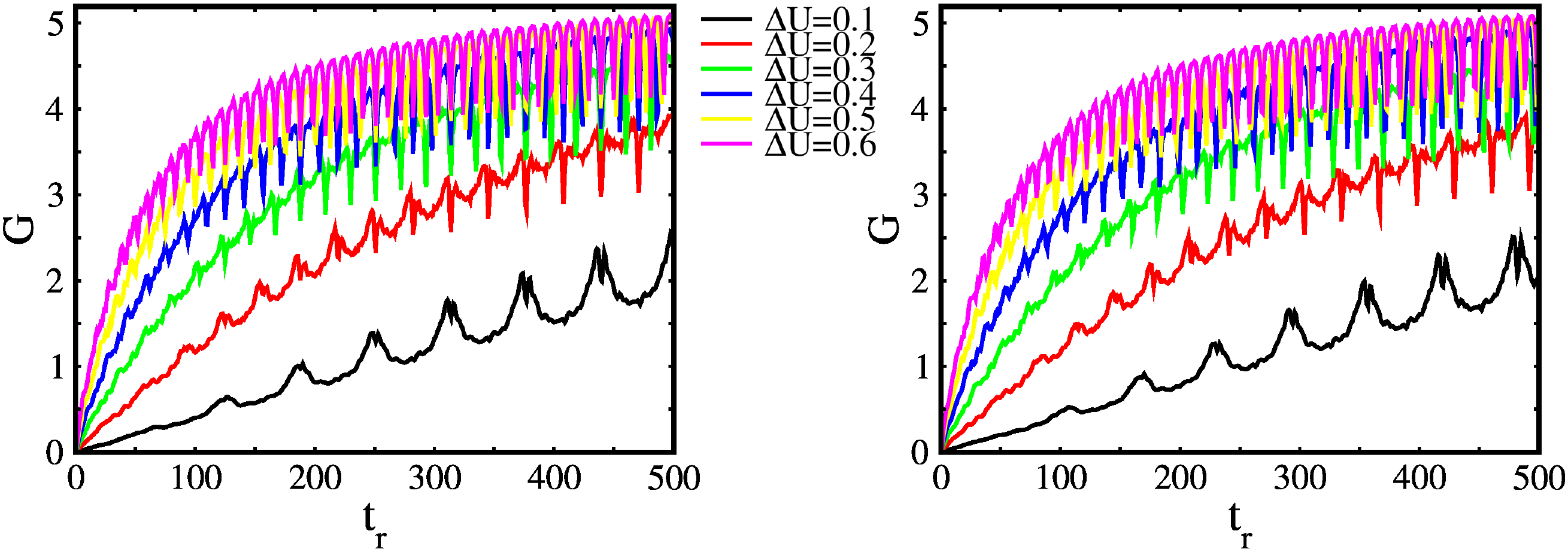}
\caption{Loschmidt echo of
entropy of entanglement at time $2t_r$ ($G=S(2t_r)$) 
as a function of $t_r$.
{\em Left panel} and {\em right panel} show the cases of $U=0$ and $U=2$ 
respectively with $\Delta U=0.1; 0.2; 0.3; 0.4; 0.5; 0.6$. 
Other parameter values are the same as in Fig.\ref{fig1}.
}
\label{fig3}
\end{center}
\end{figure}

The entropy of entanglement of the initial state is obviously 
zero $S(t=0)=0$, 
and when $U_b=U_f$ and $k_b=k_f$ the initial state
is perfectly recovered for $t=2t_r$.
We also analyze the entropy of entanglement after $2t_r$ steps given 
by $G=S(2t_r)$
as a function of $t_r$ at various
perturbation values $\Delta U=0.1,\,0.2,\,0.3,\,0.4,\,0.5,\,0.6$ 
with $U_b=U_f+\Delta U$ and $k_b=k_f=k=8$.
The results are shown  in Fig.~\ref{fig3}.
Left and right panels show the cases for $U=0$ and $U=2$ respectively.
Initially  $G(t_r)$ grows linearly with time reversal $t_r$ 
and it can be described as $G(t_r)=S(2t_r)=\alpha t_r$.
At higher times the growth is saturated
since both particles are localized in this system
(see \cite{dlstip,borgonovitip}).

\begin{figure}[h]
\begin{center}
\includegraphics[width=0.8\textwidth]{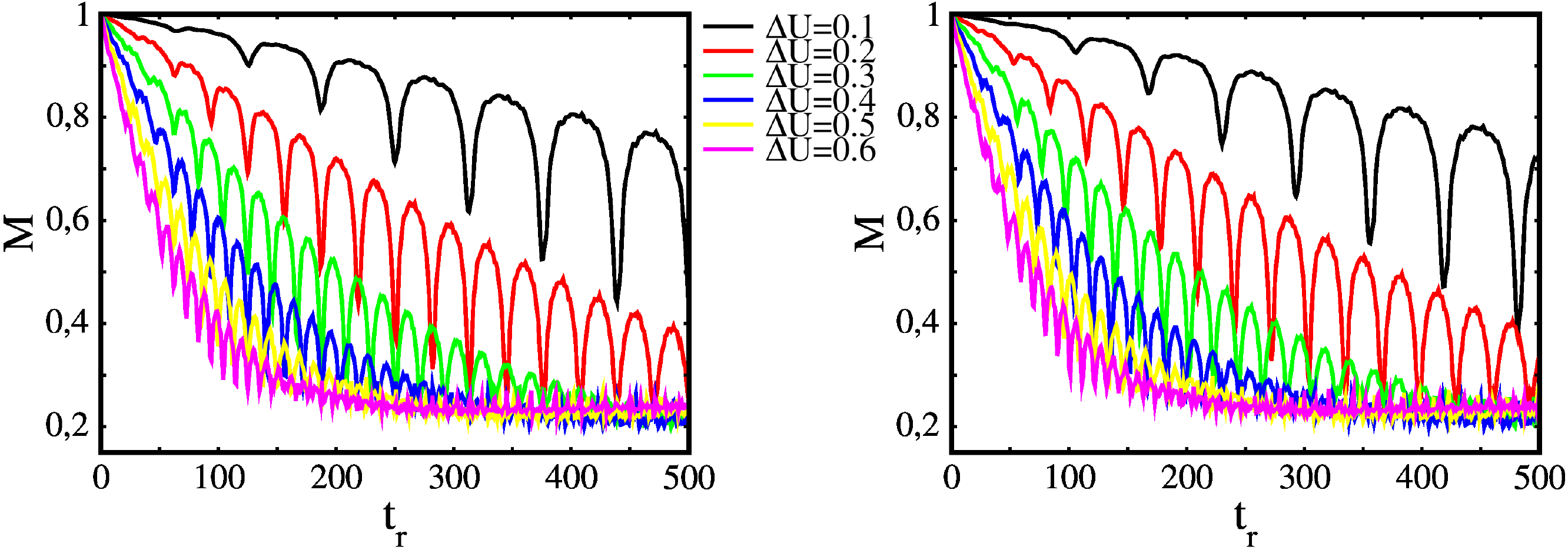}
\caption{Loschmidt echo 
$M=\vert\langle \psi (t=0)\vert\psi(t=2t_r)\rangle\vert^2$ 
as a function of $t_r$.
{\em Left panel} and {\em right panel} show the cases of $U=0$ and $U=2$ 
respectively with $\Delta U=0.1; 0.2; 0.3; 0.4; 0.5; 0.6$. 
Other parameter values are the same as in Fig.\ref{fig1}.
}
\label{fig4}
\end{center}
\end{figure}

We also study 
the usual Loschmidt echo defined as (see e.g.\cite{jalabert}):
\begin{equation}
M(t_r)=\vert \langle\psi(t=0)\vert 
U_{\rm KR}^{\dagger}(U_b,k_b)^{t_r} 
U_{\rm KR}(U_f,k_f)^{t_r}\vert\psi(t=0)\rangle \vert^2 
\end{equation}
where the initial state evolves $t_r$ steps with forward parameter 
values $U_f,k_f$ and then $t_r$ steps with backward parameter 
values $U_b,k_b$. 
Fig.~\ref{fig4} shows the Loschmidt echo as a function of time reversal $t_r$
for the same parameters of Fig.~\ref{fig1} with
$U=0$ and $U=2$ in left and right panels respectively
where $\Delta U=0.1; 0.2; 0.3; 0.4; 0.5; 0.6$ and $k_b=k_f=8$.
At short times the decrease of the echo $M(t_r)$ can be described as
the exponential decay $M(t_r) = \exp(-\Gamma t_r)$.

\begin{figure}[h]
\begin{center}
\includegraphics[width=0.8\textwidth]{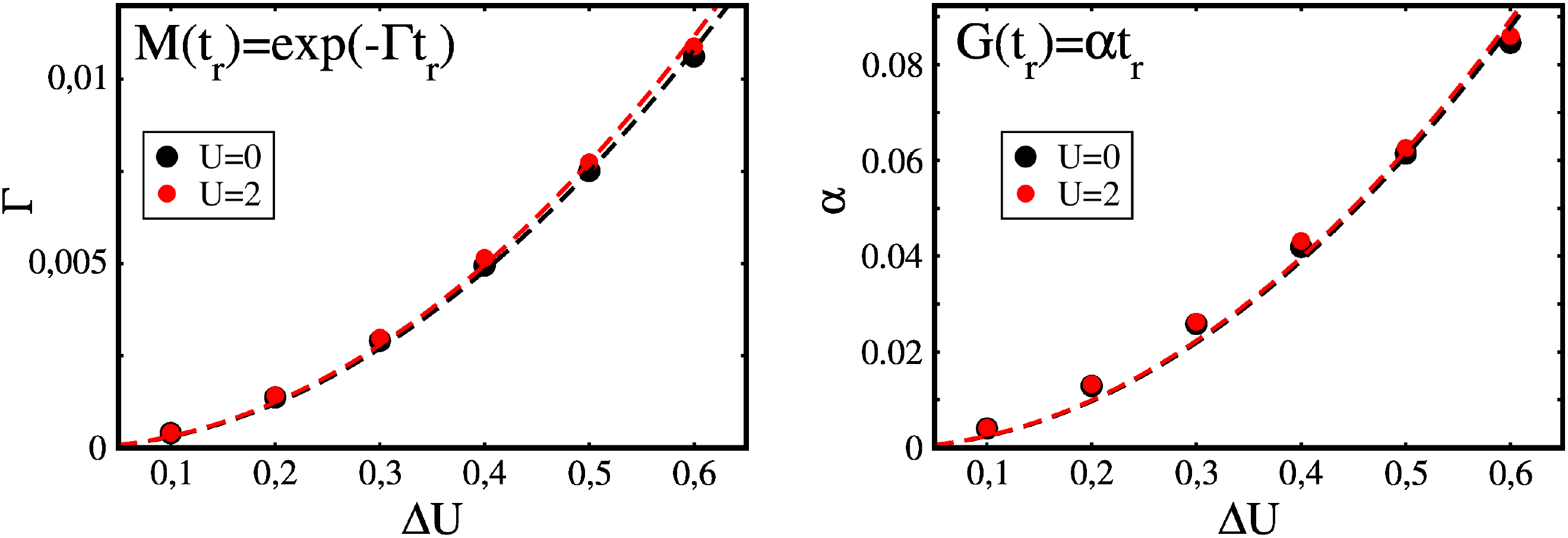}
\caption{
{\em Left:} Exponential decay rate $\Gamma$ of the Loschmidt echo vs. 
$\Delta U$ for $U=0$ (black circles) and $U=2$ (red circles);
the values of $\Gamma$ are obtained from the exponential fit 
$M(t_r)=\exp{(-\Gamma t_r)}$ for $t_r\in[0,10/\Delta U]$ 
from the curves of Fig.~\ref{fig4}.
{\em Right:}
Linear growth rate of final entropy of entanglement 
$G(t_r)=S(2t_r)$
vs. $\Delta U$ for $U=0$ (black circles) 
and $U=2$ (red circles); the values of $\alpha$ are obtained 
from the linear fit $G(t_r)=\alpha t_r$ for $t_r\in[0,10/\Delta U]$ 
from the curves of Fig.~\ref{fig3}.
The dashed curves in both
panels show the fits of $\Gamma$ and $\alpha$ by square dependence on $\Delta U$
(see text).
}
\label{fig5}
\end{center}
\end{figure}

We have fitted curves of Fig.\ref{fig3} in the interval 
$t_r\in[0,10/\Delta U]$ by the linear fit $G(t_r)=\alpha t_r$.
The obtained values of $\alpha$ 
for $U=0$, $U=2$, and $\Delta U=0.1,0.2,0.3,0.4,0.5,0.6$ 
are shown in the right panel of Fig.~\ref{fig5}.
Since the Loschmidt echo decays exponentially with time reversal $t_r$ for 
short times, 
we have fitted the curves of Fig.\ref{fig4} in the interval 
$t_r\in[0,10/\Delta U]$ with $M(t_r)=\exp{(-\Gamma t_r)}$. 
The obtained values $\Gamma$ 
for $U=0$ and $U=2$, and $\Delta U=0.1,0.2,0.3,0.4,0.5,0.6$ 
are shown in the left panel of Fig.\ref{fig5}.

The results presented in Fig.~\ref{fig5} show that the dependence of
$\Gamma$ and $\alpha$ of perturbation $\Delta U$ is well described by
the quadratic growth: $\Gamma = A (\Delta U)^2$ and
$\alpha = B  (\Delta U)^2$ with the fit values
$A =  0.030 \pm 0.001$ (for $U=0$),
$A =  0.031 \pm 0.001$ (for $U=2$);
$B=0.169 \pm 0.002$ (for $U=0$),
$B=0.172 \pm 0.002$ (for $U=2$). Such a dependence appears naturally
from the Fermi golden rule and is well known for the usual Loschmidt echo
in the case of a small parameter perturbation
(see e.g. \cite{prosen,jacquod,jalabertsch}).
For the Loschmidt echo of entanglement $G(t_r)$
the linear growth of $G$ with time $t_r$ and 
the proportionality of the growth rate $\alpha \propto (\Delta U)^2$, 
as the Fermi golden rule, is less obvious
since this implies that the number of involved states
grows exponentially with time $t_r$.
We expect that the linear time dependence
$G= \alpha t_r$ (at small times)
is in some way linked to an exponential spreading
of the wave packet with time, which results
from the positive classical Kolmogorov-Sinai entropy
$h$. In fact for the classical chaotic dynamics
$h$ gives the entropy growth rate in time \cite{arnold,sinai}.
In this context the perturbation $\Delta U$ gives
a quadratic correction to this rate.

\section{Poincar\'e recurrences of entanglement}
\label{sec4}

Here we consider the kicked rotator for two interacting particles with absorption. 
The time evolution is given by:
\begin{equation}
\label{eqtimeevolut}
\ket{\psi(t+1)}=U_a\,U_{\rm KR}\ket{\psi(t)} \; .
\end{equation}
Here, the operator $U_{\rm KR}$ is given by (\ref{eqUKR1}),
(\ref{eqUKR2}), (\ref{eqUKR3}) and 
the operator $U_a$ represents absorption
at the borders $\pm L/2=\pm N/4$, i.e.,  
it is a non-unitary operator which is diagonal in momentum representation 
and with eigenvalues being either $1$ if $-L/2\le p_1<L/2$ and 
$-L/2\le p_2<L/2$ 
or $0$ for all other cases. Due to the non-unitarity of this operator 
the survival probability $P(t)=\|\ket{\psi(t)}\|^2$ of both particles 
decays with time. In the 
following, we compute all quantities (except for $P(t)$ itself) by 
first renormalizing the state 
$\ket{\psi(t)}\to\ket{\psi(t)}/\|\ket{\psi(t)}\|$.
This model was already extensively studied for the case of one particle in 
\cite{borgonovi1,maspero1,maspero2,dlsweyl}.

We start with the initial entangled state:
\begin{equation}
\label{eqinitstate}
\ket{\psi(t=0)}=
\alpha_1(0)\ket{u_1(0)}\otimes\ket{v_1(0)}+
\alpha_2(0)\ket{u_2(0)}\otimes\ket{v_2(0)}
\end{equation}
with $\alpha_1(0)=\alpha_2(0)=1/\sqrt{2}$ and 
\begin{eqnarray}
\label{equ1}
\ket{u_1(0)}&=&\ket{p_1=6\Delta p}\\
\label{equ2}
\ket{u_2(0)}&=&\ket{p_1=7\Delta p}\\
\label{eqv1}
\ket{v_1(0)}&=&\ket{p_2=7\Delta p}\\
\label{eqv2}
\ket{v_2(0)}&=&\ket{p_2=8\Delta p}
\end{eqnarray}
where $N$ is the total system size which is chosen as a multiple of 
the minimal size 128 and $\Delta p=N/128$ is a size dependent 
scaling factor. Here the state $\ket{p_j}$ ($j=1,2$) is the momentum 
eigenstate of particle $j$ with integer (quantum) momentum values 
$-N/2\le p_j<N/2$ and $\hat p_j$ is the associated momentum operator 
of particle $j$. 
In particular, the states $\ket{u_{1,2}}$ correspond to the first 
and the states $\ket{v_{1,2}}$ correspond to the second particle.

In this and the following sections, we present in figures 
results for $U=0$ and $U=2$, $U_r=1$. For this section, 
we have also computed additional results for other three interaction cases 
$U=-2$, $U_r=1$ and $U=\pm 2$, $U_r=0$ which are qualitatively 
similar. The corresponding data are not shown in figures 
but we mention in the discussions the important differences.

Concerning the parameters $T$ and $k$ we choose:
$T=\hbar$, $k=K/\hbar$ with $K=7$ and 
$\hbar$ is fixed by the condition that $k=N/8$ implying 
$\hbar=K/k=56/N$. The value $K=7$ corresponds to strong 
chaos with two small islands and has the particular feature 
that the classical decay (in case with of absorption at some border), 
described by the statistics of Poincar\'e recurrences, is also exponential 
in contrast to a typical subtle power law decay in presence of 
significant stable islands \cite{chsh,frahmulam,chirikov2000b}. 

To compute numerically the application of the operator $U_{\rm KR}$ 
to the state $\ket{\psi(t)}$ given in momentum representation 
we apply first the diagonal phase shift due to (a half of) the 
free rotation (the term (\ref{eqUKR3})), 
transform the state to phase representation using a backward 2d-FFT, 
apply the diagonal kick operator (the term (\ref{eqUKR2})), transform the 
state back to momentum representation using a forward 2d-FFT, and 
then we apply the second half of the free rotation (the term (\ref{eqUKR1})).
In absence of interaction $U=0$, it also possible to use 1d-FFT and 
one-particle operators for free rotation and the kick operator 
applied directly to the Schmidt states $\ket{u_{1,2}}$ 
and $\ket{v_{1,2}}$ (see also \cite{eprchaos} for some technical details 
on this).

We note that in classical momentum units: $p_{cl}=\hbar p_j$ the absorption 
border is $p_{\rm max}=\hbar L/2=(56/N)\cdot N/4=14=(2\pi)\cdot 2.228$ 
corresponding to $2.228$ classical momentum cells in each direction 
(from $0$ to $\pm p_{\rm max}$). The classical momentum values of the 
two states  $\ket{u_2}$ and $\ket{v_1}$, given 
in (\ref{equ2}) and (\ref{eqv1}) and used in the initial 
state (\ref{eqinitstate}), correspond to:
$\hbar\,(7N/128)=(56/N)\,(7N/128)=49/16=(2\pi)\cdot 0.4874\approx (2\pi)\cdot 0.5$, 
which is in the middle of the classical cell in order to avoid 
the two small stable islands at $p_{cl}=0$ (and 
some non-trivial phase values). The other two states 
$\ket{u_1}$ and $\ket{v_2}$ are close by to $\ket{u_2}$ or $\ket{v_1}$ 
by a factor of $6/7$ or $8/7$ respectively.

Assuming a simple 1d classical diffusive process 
\begin{equation}
\label{eqDiffus}
\frac{\partial \rho_{cl}(p,t)}{\partial t}=D\,
\frac{\partial^2\rho_{cl}(p,t)}{\partial p^2}
\end{equation}
for the classical density $\rho_{cl}(p)$ (of one particle), 
with diffusion 
constant $D=k^2/4=N^2/256$ and 
absorption at $p=\pm L/2=\pm N/4$, i.e., Dirichlet boundary conditions  
$\rho_{cl}(\pm L/2)=0$ at the absorption border, 
one can show that the classical survival probability 
(for one particle) decays (for long times) as $P_1(t)\sim\exp(-t/t_{\rm Th})$ 
with $t_{\rm Th}=L^2/(\pi^2 D)=64/\pi^2\approx 6.4846$ 
being the Thouless time. 
For both particles we expect $P(t)=P_1^2(t)$ such that 
$P(t)^{1/2}\sim\exp(-t/t_{\rm Th})$. 

When applying the quantum iteration (\ref{eqtimeevolut}), the state 
at a given (integer) iteration time $t$ 
takes the more general form (after renormalization)~:
\begin{equation}
\label{eqschmidtgeneral}
\ket{\psi(t)}=\sum_{i=1}^{L} \alpha_i(t)\ket{u_i(t)}\otimes\ket{v_i(t)}\ .
\end{equation}
We describe in Appendix A.1 some details on the practical computation 
of the general Schmidt decomposition and also some other related properties. 
Furthermore, we also show in Appendix A.2, 
that in absence of interaction there is indeed a more efficient numerical 
scheme to recompute the Schmidt decomposition and in particular 
at arbitrary time $t$ there are only 
two non-vanishing singular values being $\alpha_1$ and $\alpha_2$ and 
$\alpha_i=0$ for $i=3,\ldots,L$ (note that the effective matrix 
size of $\psi(p_1,p_2)$ is $L\times L$ with $L=N/2$ due to the absorption
process). 

Using the singular values one can compute the 
entropy of entanglement defined as \cite{chuang}:
\begin{equation}
\label{eqentropydef}
S(t)=-\mbox{Tr}[\rho_j\log_2(\rho_j)]
=-\sum_{i=1}^L \alpha_i^2(t)\,\log_2[\alpha_i^2(t)]
\end{equation}
(for either $j=1$ or $j=2$).
In this definition we use the logarithm 
with respect to the base $2$ such that the entropy of the initial state 
(\ref{eqinitstate}) is exactly $S(t=0)=1$. 

For $U=0$, we stop the iteration 
when $\ket{u_1}$ converges (up to a global phase factor) 
with an error below $10^{-12}$ (for the norm of the difference 
vector with proper phase) between two time steps 
(and then we also iterate up to the next power of two 
for the iteration time). 

For $U\neq 0$, we stop the quantum interaction when the change of 
entropy between two time steps is below $10^{-14}$ (and then 
we iterate up to the next power of two for the iteration time). 
It turns that for the two cases with $U_r=1$
also the state $\ket{\psi(t)}$ converges up to a global phase factor
to a limit state $\ket{\psi_{\infty}}$ 
, i.e., $\ket{\psi(t)}\to e^{i\phi(t)}\,\ket{\psi_{\infty}}$ for 
$t\to\infty$ with a time 
dependent phase factor. This is coherent with the picture that 
the non-unitary iteration operator $U_a\,U_{\rm KR}$ has complex 
eigenvalues of the form $\exp(-\gamma_i/2+i\omega_i)$ with 
$0<\gamma_1<\gamma_2<\ldots$ and that 
$\ket{\psi_{\infty}}$ is the eigenvector of the leading mode with 
absorption rate $\gamma_1$ and furthermore the time dependent phase 
factor is $\phi(t)=\omega_1\,t+$const. 
For $U_r=1$ this convergence is rather fast and comparable in speed to 
the convergence of the entropy. However, for $U_r=0$ this convergence 
to the limit state is quite slower. 
We have also verified that for $U=0$ the general method of computing 
all singular values, suitable for arbitrary $U$ values, 
produces the same results (up to numerical precision) as the more efficient 
method explained in Appendix A.2.

\begin{figure}
\begin{center}
\includegraphics[width=0.8\textwidth]{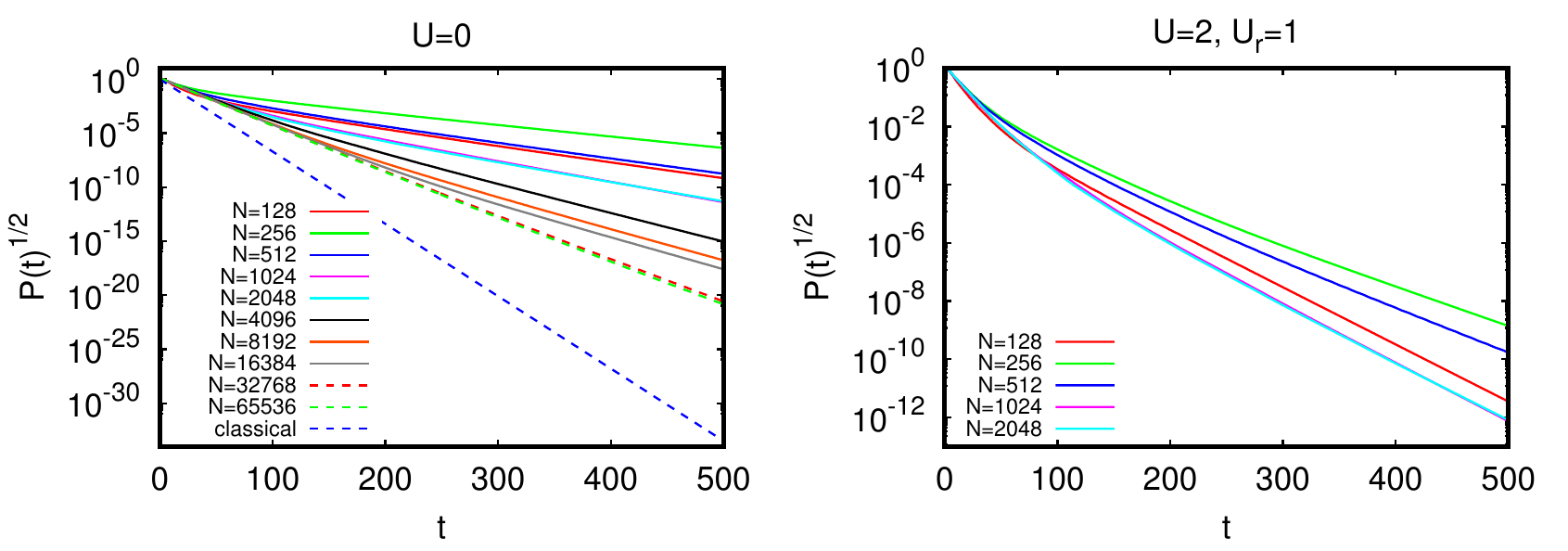}
\caption{
{\em Left:} Decay of $P(t)^{1/2}=\||\psi(t)\rangle\|$ where the norm 
is obtained before renormalization of the state  $|\psi(t)\rangle$
for interaction value $U=0$ and system size  $128\le N\le 65536$.
The dashed blue line shows the classical decay corresponding  to 
$P(t)^{1/2}=\exp(-t/t_{\rm Th})$ (for two particles) and 
with $t_{\rm Th}=64/\pi^2\approx 6.4846$ being the Thouless time.
{\em Right:} Decay of $P(t)^{1/2}=\||\psi(t)\rangle\|$ 
for interaction values $U=2$, $U_r=1$ and system size $128\le N\le 2048$.
The curves for $N=1024$ (pink) and $N=2048$ (cyan) are very close. 
}
\label{fig6}
\end{center}
\end{figure}

Fig.~\ref{fig6} shows the quantity $P(t)^{1/2}$ obtained from the 
quantum iteration (\ref{eqtimeevolut}) for the two cases 
$U=0$ with system size $128\le N\le 65536$ (left panel) 
and the interaction case 
$U=2$, $U_r=1$ with system size $128\le N\le 2048$ (right panel). 
In the left panel also the classical decay based on the model of 
simple classical diffusion is shown. For $U=0$ the quantum decay is for 
long time scales strongly reduced with respect to the classical decay 
and this effect is strongest for small values of $N$. For larger values 
of $N$ the time scale where the semiclassical limit is valid increases as 
can be seen from the two curves for $N=32768$ and $N=65536$. 
However, even for this semiclassical quantum decay the exponential decay time 
is by a factor $\sim 1.6$ 
longer than the classical Thouless time. This can be explained by the 
fact that 
the simple diffusive model is not very accurate on a quantitative level 
due to the small number of classical cells ($\sim 4.5$ in total). 
In this case, despite the strong chaos for $K=7$, the effects of the 
classical phase space structure may cause significant deviations with 
respect to the simple diffusive model. 
For the interaction case, the decay visible for the more modest system sizes 
up to $N=2048$ is comparable to the decay for $U=0$ at same system size values.
This also holds for the other interaction cases not shown in the figure.

\begin{figure}
\begin{center}
\includegraphics[width=0.8\textwidth]{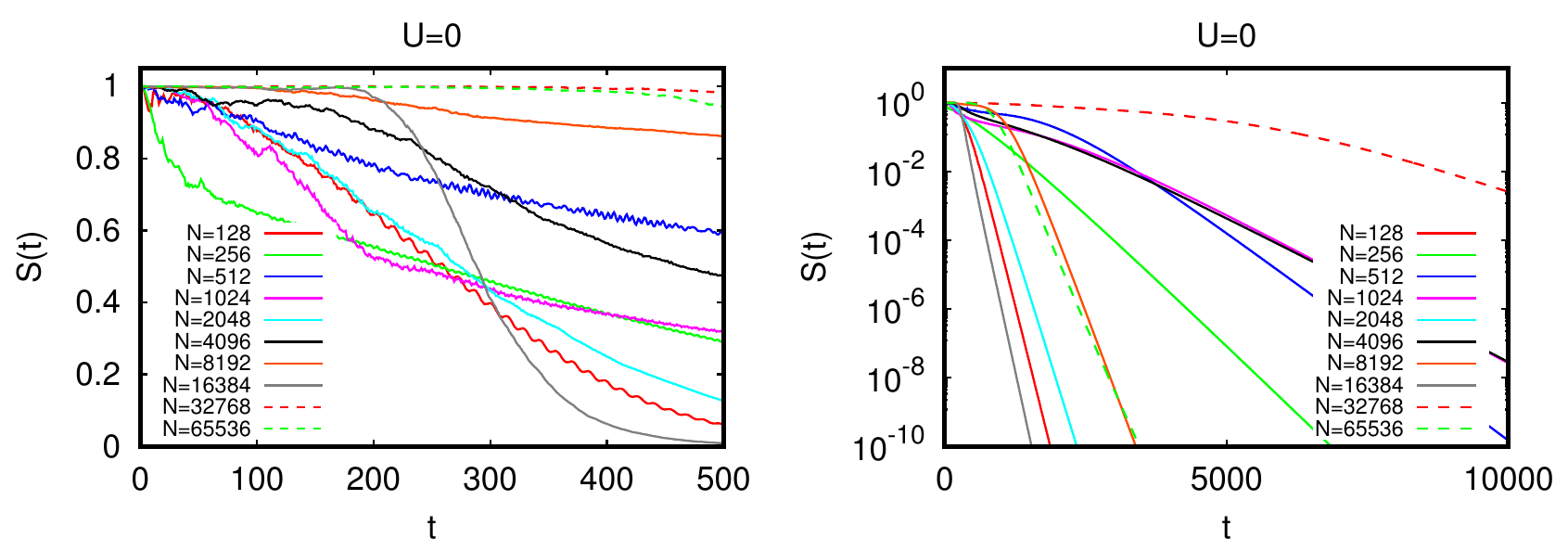}
\caption{
{\em Left:} Decay of the entropy of entanglement $S(t)$
for interaction value $U=0$ and system size  $128\le N\le 65536$.
{\em Right:} As left panel but using a logarithmic scale for the entropy 
and an increased time range. 
}
\label{fig7}
\end{center}
\end{figure}

Fig.~\ref{fig7} shows for $U=0$ and different values of $N$ 
the decay of the entropy of entanglement $S(t)$ which converges to 0 
for $t\to\infty$ for all cases. However, the exponential 
decay rates for long time scales vary strongly with system size 
in a non-systematic way, e.g., the long time decay for $N=32768$ 
is significantly slower than for the other cases. 
Furthermore, even for a case with strong long time decay, the onset 
of this decay may be quite late, e.g., for $N=65536$ the entropy 
is rather constant close to unity for $t\le 500$ with values 
clearly above the data for most of the other cases and 
starts to decay quite strongly at $t\approx 1000$ with final values 
clearly below the values for several of the other cases. 

In particular, 
we find that $\alpha_1(t)\to 1$ and $\alpha_2(t)\to 0$ for 
$t\to\infty$ and (up to phase factors) $\ket{v_1(t)}=\ket{u_1(t)}$ and 
$\ket{v_2(t)}=\ket{u_2(t)}$. (See below for the corresponding Husimi 
functions for some examples.) In absence of interaction the non-unitary 
iteration operator is simply a tensor product of two independent 
(and identical) one-particle non-unitary iteration operators:
$U_a\,U_{\rm KR}=(U_a^{(1)}\,U_{\rm KR}^{(1)})\otimes 
(U_a^{(2)}\,U_{\rm KR}^{(2)})$.
During the time iteration both $\ket{u_1(t)}$ and $\ket{v_1(t)}$ 
converge to the leading eigenvector of the one-particle non-unitary iteration 
operator and the long time decay of the survival probability 
(of both particles) is 
simply $P(t)\sim \exp(-2\gamma_1^{(1)}\,t)$ where $\gamma_1^{(1)}$ 
is the absorption rate of this leading eigenvector.
However, after global renormalization, 
the entropy is determined by $\alpha_2$ via 
\begin{equation}
\label{eqSsmallalpha}
S(t)
\approx -\alpha_2^2\log_2(\alpha_2^2)-\alpha_1^2\log_2(\alpha_1^2)
\approx -\alpha_2^2\log_2(\alpha_2^2)+\alpha_2^2/\ln(2)
\end{equation}
for $\alpha_2\ll 1$ since $\alpha_1^2=1-\alpha_2^2$. 
We expect that $\alpha_2\sim \exp[-(\gamma_2^{(1)}-\gamma_1^{(1)})\,t]$
where $\gamma_2^{(1)}$ is the absorption rate of the second 
eigenvector (of the non-unitary one-particle iteration operator) 
which is $\sim \ket{v_2(t)}=\ket{u_2(t)}$ (up to a phase factor) 
for $t\to\infty$. Therefore, the entropy should decay as:
\begin{equation}
\label{eqSdecayU0}
S(t)\sim \left((\gamma_2^{(1)}-\gamma_1^{(1)})\,t+{\rm const.}\right)\,
\exp[-2(\gamma_2^{(1)}-\gamma_1^{(1)})\,t]\ .
\end{equation}
For $U=0$ and $N=1024$, we obtain from exponential fits of 
$P(t)$ and $\alpha_2(t)$ that 
$\gamma_1^{(1)}=0.0421459$ and $\gamma_2^{(1)}-\gamma_1^{(1)}=0.00106019$ 
(with virtually no statistical fit error for both quantities) 
and also (\ref{eqSdecayU0}) 
is numerically very well confirmed. The inverse of 
$\gamma_1^{(1)}$ provides the decay time $t_q\approx 23.73$ which is about 
$3.66$ times larger than the classical decay time $t_{\rm Th}=64/\pi^2=6.4846$ 
(see also pink full line and blue dashed line in the 
left panel of Fig.~\ref{fig6}). 

\begin{figure}
\begin{center}
\includegraphics[width=0.8\textwidth]{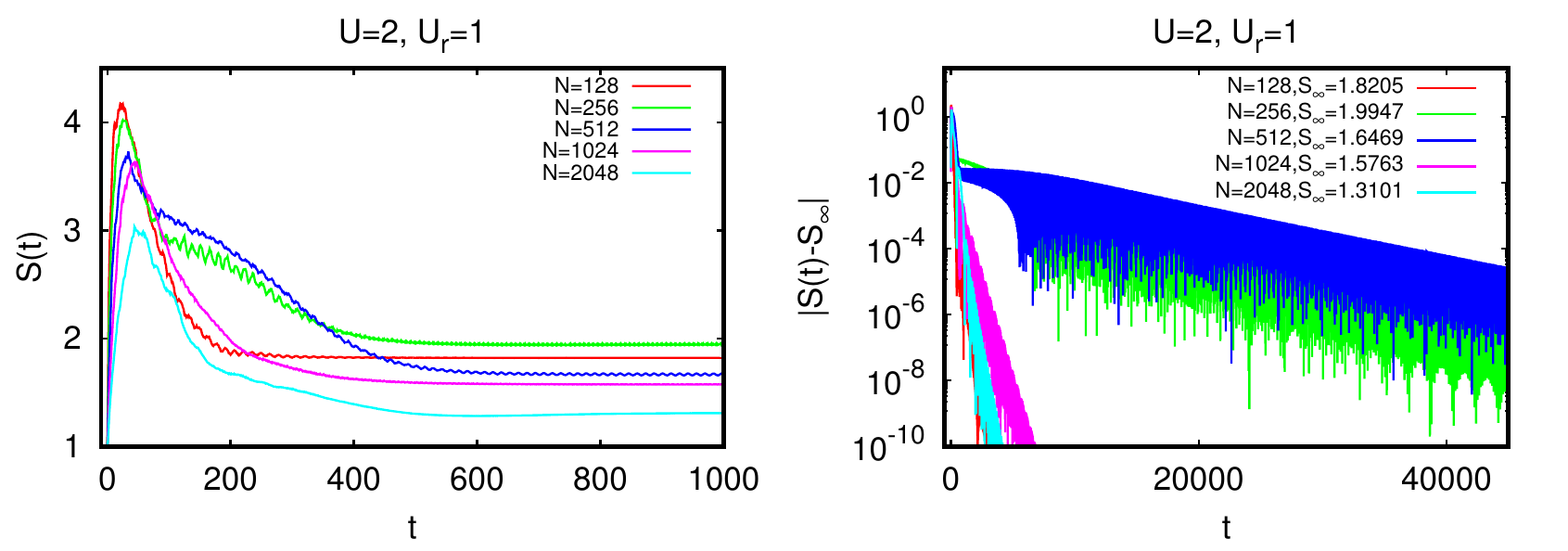}
\caption{
{\em Left:} Entropy of entanglement $S(t)$ versus iteration time 
for interaction values $U=2$, $U_r=1$ and system size  $128\le N\le 2048$.
{\em Right:} Decay of the difference $|S(t)-S_{\infty}|$ for the same values 
of interaction and system size with $S_{\infty}=\lim_{t\to\infty} S(t)$. 
The shown time scale is increased 
and a logarithmic scale for the vertical axis is used. 
The modulus of the difference is taken because $S(t)$ is for certain 
time values $t$ below the limit value due to small 
amplitude oscillations for long time scales. 
The value of $S_\infty$ is significantly larger than unity. 
}
\label{fig8}
\end{center}
\end{figure}

Fig.~\ref{fig8} shows the behavior of the entropy $S(t)$ for the 
case of $U=2$, $U_r=1$ and system size $128\le N\le 2048$.
Initially the entropy increases to maximum values between 
$S_{\rm max}\approx 4.2$ at $t\approx 25$ (for $N=128$) and 
$S_{\rm max}\approx 3.0$ at $t\approx 50$ (for $N=2048$) 
and then it decays for $t\to\infty$ exponentially
to a limit value $S_{\infty}$ which is significantly larger than 
unity and slightly below 2. The values of $S_{\infty}$ for 
each case of $N$ are also given in the right panel 
of Fig.~\ref{fig8} which shows the (modulus of the) 
difference $S(t)-S_{\infty}$ 
versus time on a longer time scale and logarithmic scale for the 
vertical axis. One also observes that for long time scales there are 
small amplitude oscillations of $S(t)$ around $S_{\infty}$.

The behavior for the case $U=-2$, $U_r=1$ (not shown in the figure) 
is qualitatively
similar with an initial increase of $S(t)$ and similar maximum values and 
with slightly different values of $S_{\infty}$. However, 
the decays for the two cases of $N=256$ and 
$N=512$ are significantly faster 
with a convergence time scale for $S(t)$ at 
$t_{\rm max}\approx 9000$ or $3000$ respectively (the other two cases 
$N=128$ and $N=1024$ have similar values of the convergence time as for 
$U=2$, $U_r=1$ and visible in Fig.~\ref{fig8}).

For the two cases of $U=\pm 2$, $U_r=0$ (not shown in the figure) 
the limit value is precisely 
$S_{\infty}=1$ and it turns out that the limit state has only two non-vanishing 
singular values $\alpha_1=\alpha_2=1/\sqrt{2}$ while $\alpha_i=0$ for 
$i\ge 3$. However, also here the entropy initially increases
to the maximum values between 
$S_{\rm max}\approx 3.5$ at $t\approx 25$ (for $N=128$) and 
$S_{\rm max}\approx 2.5$ at $t\approx 25$ (for $N=512$)
and with $S_{\rm max}\approx 2.6$ at $t\approx 50$ (for $N=1024$).
At these intermediate iteration times also the singular values 
$\alpha_i$ for $i\ge 3$ provide significant contributions to the 
state $\ket{\psi(t)}$. 
The decay of $S(t)-S_{\infty}$ for $t\to\infty$ is here very close 
to a pure exponential decay without any amplitude oscillations 
(except for $N=1024$ with some deviations from the pure exponential 
behavior) 
and with a convergence time scale of $t_{\rm max}$ between 300 (for 
$N=128$) and $t_{\rm max}$ between 2400 (for $N=1024$) 
such that $S(t)-S_{\infty}<10^{-10}$ for $t>t_{\rm max}$. 

\begin{figure}
\begin{center}
\includegraphics[width=0.8\textwidth]{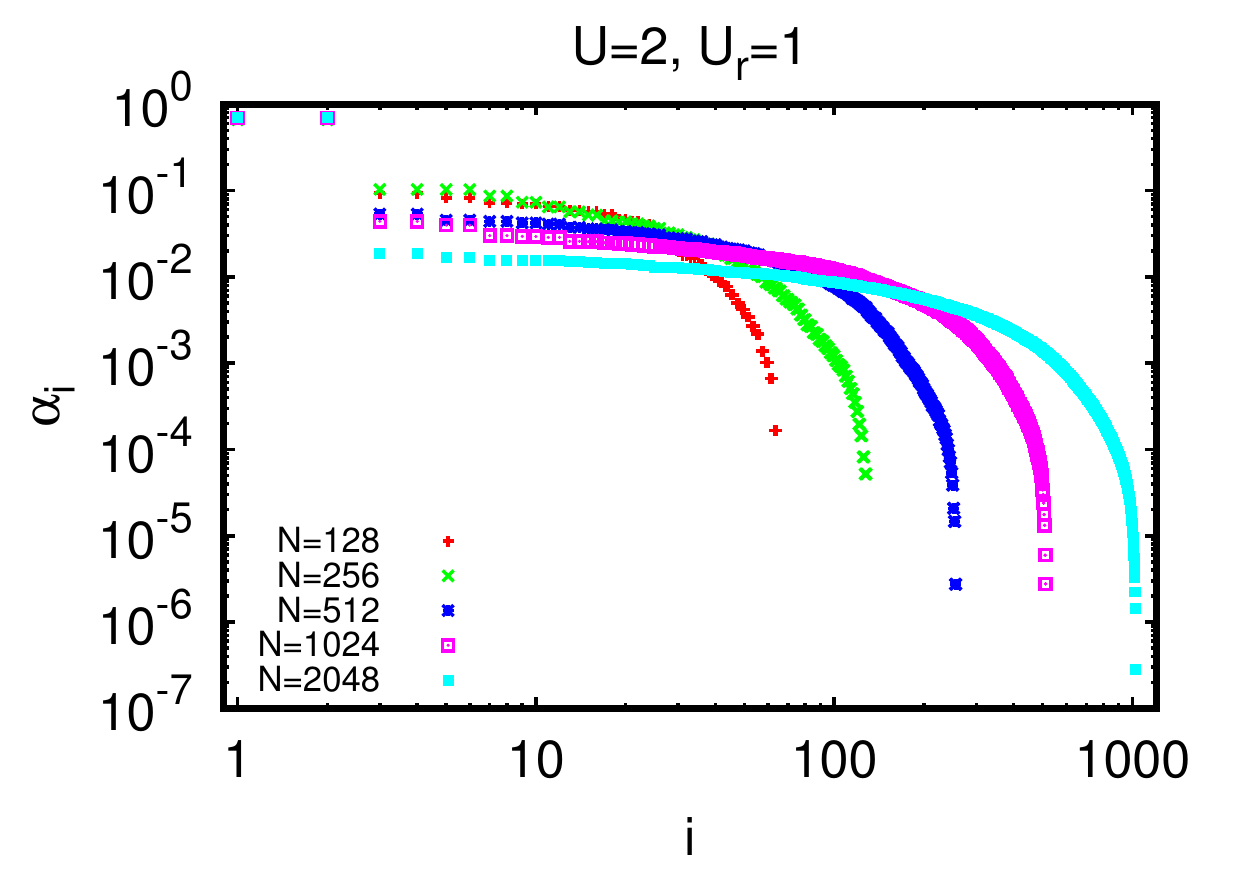}
\caption{
Singular values $\alpha_i$ (appearing in the  Schmidt decomposition of 
the limit state $\lim_{t\to\infty}|\psi(t)\rangle$) versus index $i$
for interaction values $U=2$, $U_r=1$ and system size $128\le N\le 2048$.
Both axis are shown on a logarithmic scale.
The singular values appear in degenerate pairs since the limit state is 
anti-symmetric with respect to particle exchange. The values of the top pair 
of singular values are $\alpha_1=\alpha_2=
0.67152,\,0.66317,\,0.68510,\,0.68965,\,0.69924 $ 
for $N=128,\,256,\,512,\,1024,\,2048$ respectively.
}
\label{fig9}
\end{center}
\end{figure}

For $U=\pm 2$, $U_r=1$ all singular values contribute to the limit state 
as can be seen in Fig.~\ref{fig9} (for $U=2$, $U_r=1$; the other case 
$U=-2$, $U_r=1$ being very similar). However the first two singular 
values $\alpha_1=\alpha_2\approx 0.66$--$0.69$ (see caption of Fig.~\ref{fig9}
for more precise values) are dominant while 
$\alpha_i\le 0.1$ for $i\ge 3$. Furthermore the singular values 
of the limit state appear in degenerate pairs. This is due to the fact that 
for all four interaction cases the limit state is anti-symmetric with 
respect to particle exchange and a general property for the singular 
values of a complex skew-symmetric matrix \cite{rogerhorn} 
(see also Appendix A.1 for some explanations on this).

We show in Supplementary Material (SupMat) that
the weight of the symmetric component of the wavefunction decays 
exponentially to zero (see Fig.~S1) for long time scales.

\begin{figure}
\begin{center}
\includegraphics[width=0.8\textwidth]{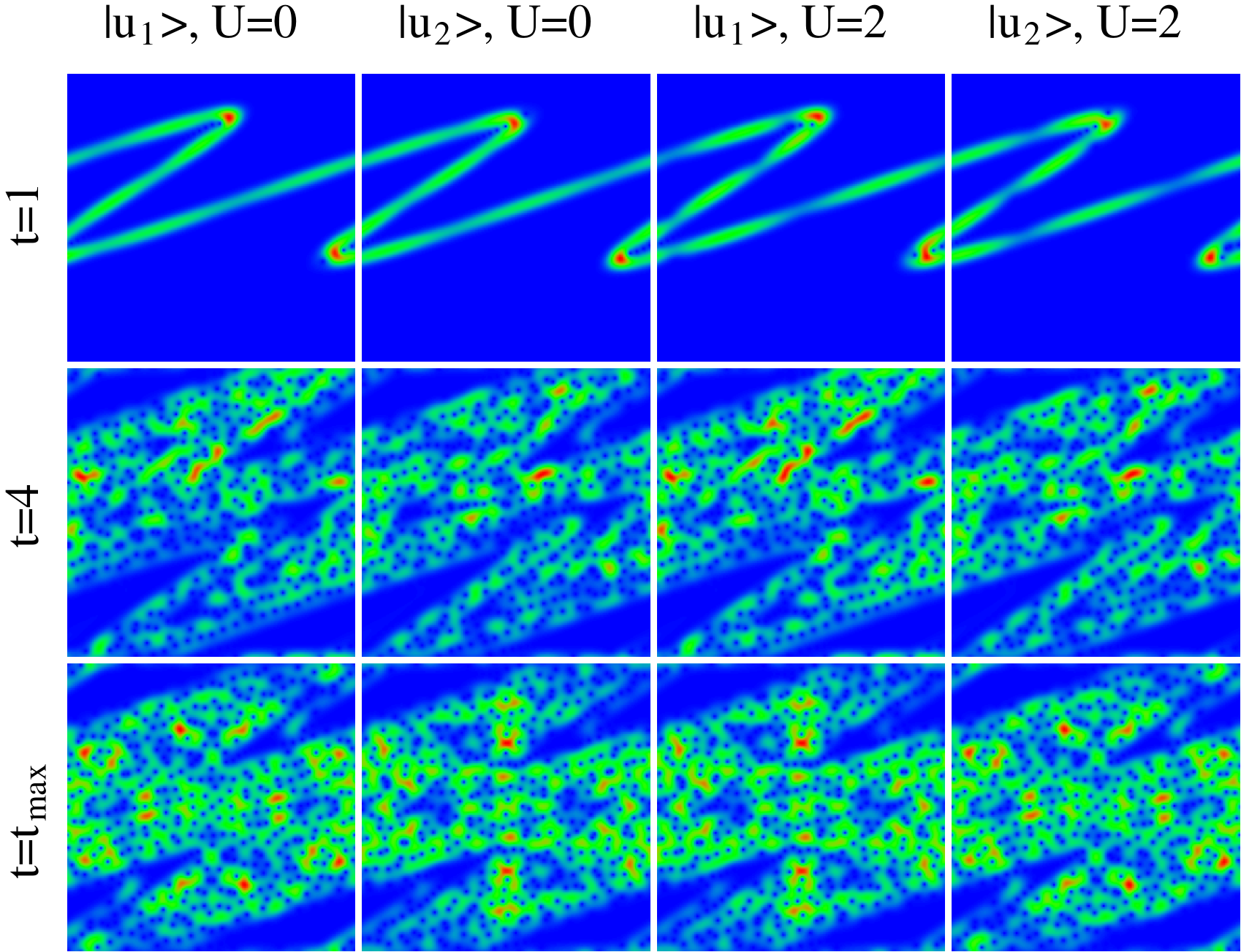}
\caption{
Husimi functions of the Schmidt states $|u_1\rangle$, $|u_2\rangle$, 
for $U=0$ or $U=2$ (with $U_r=1$)  
and $N=1024$ at three iteration times $t=1$, $t=4$ and 
$t=t_{\rm max}$ with $t_{\rm max}=16384$ for $U=0$ or $t_{\rm max}=4096$ 
for $U=2$. 
The horizontal axis corresponds to the phase  $\theta\in[-\pi,\pi]$ and 
the vertical axis to the momentum $p_{\rm cl}\in[-p_{\rm max},p_{\rm max}]$ 
where $p_{\rm max}=(2\pi)\cdot 2.228$ is the momentum absorption border 
in classical units. The colors red/green/blue correspond to 
maximal/medium/minimal values of the Husimi function. 
}
\label{fig10}
\end{center}
\end{figure}

We have also computed for $U=0$ and the interaction cases the first 
Schmidt vectors $\ket{u_{1,2}}$, $\ket{v_{1,2}}$ and the associated 
Husimi functions (see e.g. \cite{husimi,qcfrahm,eprchaos} 
for details on the definition and computation of Husimi functions). 
The vectors $\ket{u_{1,2}}$ for $U=0$ and 
$U=2$, $U_r=1$ (all for $N=1024$) 
are shown in Fig.~\ref{fig10} for three time values 
being $t=1$, $t=4$ and some very long time being $t=t_{\rm max}=16384$ 
for $U=0$ or $t=t_{\rm max}=4096$ for $U=2$, $U_r=1$. 
For $t=1$, these states occupy the same manifold but for the two 
interaction cases the densities at some positions on 
this manifold are reduced in comparison to the non-interacting case.

For $t=4$ the phase space structure is quite complicated but rather 
similar between the interacting and non-interacting cases for both 
states $\ket{u_{1,2}}$.
At long times the state $\ket{u_1}$ of $U=0$ is very close 
to the state $\ket{u_2}$ of $U=2$, $U_r$ and similarly between 
$\ket{u_2}$ of $U=0$ and $\ket{u_1}$ of $U=2$, $U_r=1$. 

The states $\ket{v_{1,2}}$ are not shown but for long times they 
are determined by $\ket{u_{1,2}}$ depending on the interaction:
for $U=0$ they are given by~: $\ket{u_1}=\ket{v_1}$ and $\ket{u_2}=\ket{v_1}$
(up to phase factors), or more explicitly:
\begin{equation}
\label{eqpsilimU0}
\ket{\psi_{\infty}}=C_1 \alpha_1\ket{u_1}\otimes\ket{u_1}
+ C_2 \alpha_2\ket{u_2}\otimes\ket{u_2}
\end{equation}
where $\alpha_1\approx 1$, $\alpha_2=\sqrt{1-\alpha_1^2}\ll 1$ 
and $C_1$, $C_2$ represent unknown (and time dependent) phase factors. 

For $U=\pm 2$ and $U_r=0,1$ we have a different situation where 
(up to a global phase factor) $\ket{v_1}=\ket{u_2}$ and 
$\ket{v_2}=-\ket{u_1}$ which is confirmed by the Husimi functions of these 
states and the sign is due to the established anti-symmetry of the limit 
state,  or more explicitly:
\begin{equation}
\label{eqpsilimU2}
\ket{\psi_{\infty}}=C \alpha_1
\Bigl(\ket{u_1}\otimes\ket{u_2}-\ket{u_2}\otimes\ket{u_1}\Bigr)
+\alpha_3(\cdots)+\ldots
\end{equation}
where $\alpha_1=\alpha_2\approx 0.7$, $C$ is an unknown phase factor 
and (for $U_r=1$) there are also smaller contributions due to 
$\alpha_i$ for $i\ge 3$.

As already discussed above, the limit state (\ref{eqpsilimU0}) 
for $U=0$ has a decay rate $\gamma_1^{(1)}=0.0421459$ (for $N=1024$) 
since both modes $\ket{u_1}=\ket{v_1}$ in the first term have 
the decay rate $\gamma_1^{(1)}/2$
(of the non-interacting one-particle iteration operator).
For the anti-symmetric state (\ref{eqpsilimU2}) in presence of interaction, 
neglecting the contributions of $\alpha_i$ for $i\ge 3$
and assuming that $\ket{u_{1,2}}$ represent well the first two 
decay modes of the non-interacting one-particle iteration operator, 
one would expect a decay rate of 
$(\gamma_1^{(1)}+\gamma_2^{(1)})/2=0.0426760$ where 
$\gamma_2^{(1)}$ is extracted from the difference 
$\gamma_2^{(1)}-\gamma_1^{(1)}=0.00106019$ which is according 
to the discussion above just the decay rate of $\alpha_2(t)$ at $U=0$. 

Numerically, we find that the decay rates of the limit state in presence of 
interaction (for $U_r=1$) are $\gamma_1^{(U=2,U_r=1)}=0.0455464$ and 
$\gamma_1^{(U=-2,U_r=1)}=0.045960$. Even though these 
values are quite close to the above expectation the difference is 
still 2 or 3 times larger than $\gamma_2^{(1)}-\gamma_1^{(1)}$, indicating 
that the interaction has a significant influence on the decay rate as such 
(in addition to ``imposing'' the anti-symmetric limit 
state (\ref{eqpsilimU2})). This is also plausible since for $U_r=1$ the 
contributions of the other singular values $\alpha_i$ for $i\ge 3$ 
are still quite important which can be seen from 
the limit value of $S_\infty$ which is closer to 2 than to 1.

However, for the short range interaction $U_r=0$ we have:
$\gamma_1^{(U=\pm 2,U_r=0)}=0.04264947$ 
which is indeed very close to the theoretical expectation 
(difference being the fraction $1/40$ of $\gamma_2^{(1)}-\gamma_1^{(1)}$). 
For this case, the interaction is responsible for the anti-symmetric 
combination of the first two terms in 
(\ref{eqpsilimU2}) but has otherwise no strong effect on the decay. 
This is also coherent with the fact that for this case 
$\alpha_i=0$ for $i\ge 3$ (for the limit state 
at $t\to\infty$).

One should note that for pairwise degenerate singular values the Schmidt 
decomposition is not unique and does not change if one applies 
to $\ket{u_{1,2}}$ and $\ket{v_{1,2}}$ an arbitrary $2\times 2$ 
unitary rotation (same rotation for both pairs). 
Therefore, the numerical procedure that computes 
the Schmidt vectors selects in some random way the precise choice 
of these vectors. However, despite this degree of liberty we 
have been able to verify 
that essentially the vectors $\ket{u_{1,2}}$ in the limit 
$t\to\infty$ for $U=0$ coincide roughly 
with the first two Schmidt vectors for the four interaction cases.
For $U_r=0$ there are no other Schmidt components but for $U_r=1$ there 
are with a modest weight further Schmidt components due to the interaction
which also provide a quite significant influence on the decay rate.
Apparently, the interaction imposes essentially an anti-symmetric limit 
state where both particles occupy the first two non-interacting 
absorption modes in an anti-symmetric entangled combination. However, 
for $U_r=1$ the other Schmidt components also have a significant weight 
in the entropy of entanglement which is closer to 2 than to 1.

\begin{figure}
\begin{center}
\includegraphics[width=0.8\textwidth]{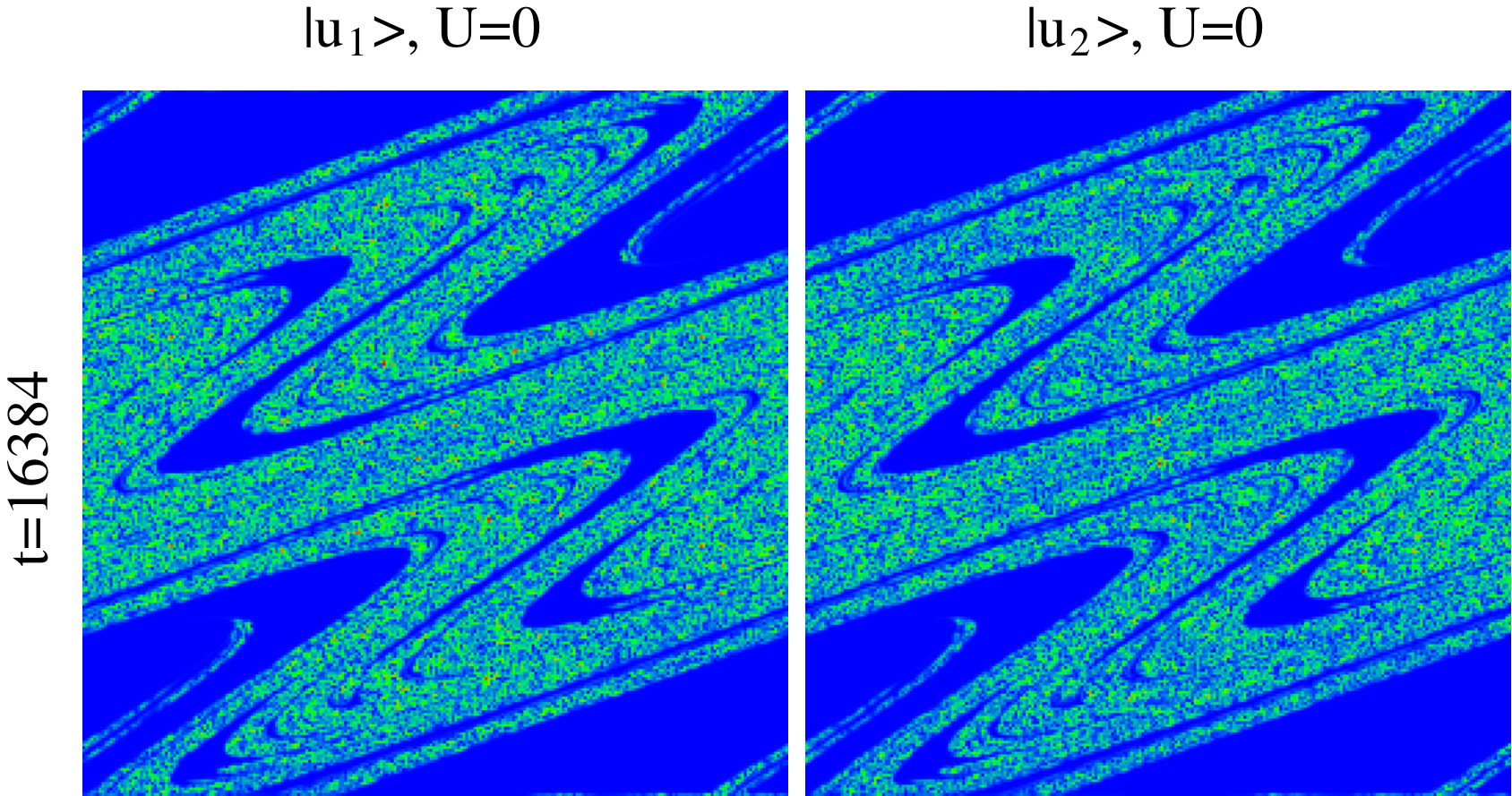}
\caption{
Husimi functions of the Schmidt states $|u_1\rangle$, $|u_2\rangle$, 
for $U=0$ and $N=65536$ at final iteration time $t=16384$. The signification 
of both axes and the color codes is the same as in Fig.~\ref{fig10}.
}
\label{fig11}
\end{center}
\end{figure}

In Fig.~\ref{fig11}, we also show the Husimi functions of $\ket{u_{1,2}}$ 
for $U=0$ and a larger system size $N=65536$ at final iteration time $t=16384$.
The overall phase space structure of these modes is rather complicated 
but there are some positions, with very small red dots, where the 
density is locally enhanced. The Husimi function has a fractal structure
corresponding to a fractal repeller of non-escaping classical orbits
(compare with the one-particle case discussed in \cite{dlsweyl}).

\section{Poincar\'e recurrences of entanglement with absorption of one particle }

In this Section, we consider a different case with the same time 
evolution as in the previous Section (exact same parameters; in 
particular $K=7$) 
but the absorption at the absorption border is applied only to the second 
particle while the first particle can move in the full available 
phase space (with periodic boundary conditions) and is never absorbed.
We call this the asymmetric absorption case
(in contrast to the symmetric absorption case
of the previous Section).

\begin{figure}
\begin{center}
\includegraphics[width=0.8\textwidth]{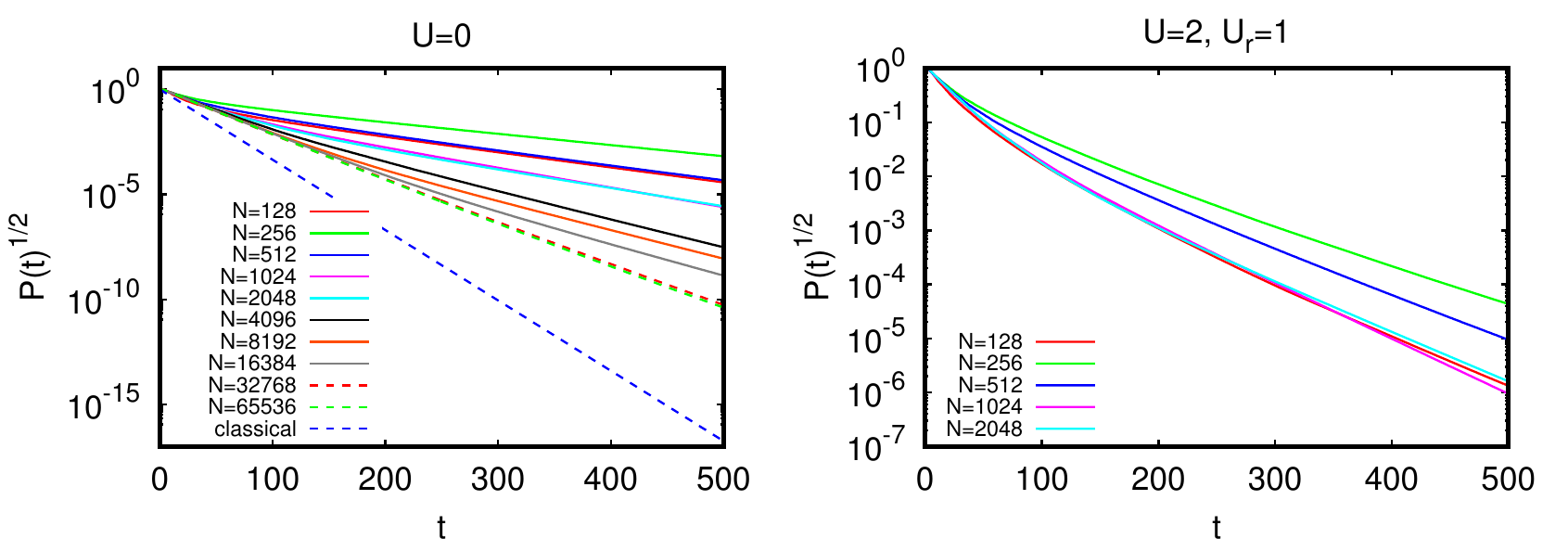}
\caption{
{\em Left:} Decay of $P(t)^{1/2}=\||\psi(t)\rangle\|$ where the norm 
is obtained before renormalization of the state  $|\psi(t)\rangle$
for interaction value $U=0$, system size  $128\le N\le 65536$
and absorption only for the second particle. 
The dashed blue line shows the classical decay corresponding  to 
$P(t)^{1/2}=\exp(-t/(2\,t_{\rm Th}))$ (for {\em one} particle) and 
with $t_{\rm Th}=64/\pi^2\approx 6.4846$ being the Thouless time.
{\em Right:} Decay of $P(t)^{1/2}=\||\psi(t)\rangle\|$ 
for interaction values $U=2$, $U_r=1$ and system size $128\le N\le 2048$.
}
\label{fig12}
\end{center}
\end{figure}

Fig.~\ref{fig12} shows for this case 
the quantity $P(t)^{1/2}$ obtained from the 
quantum iteration (\ref{eqtimeevolut}) for 
$U=0$ with system size $128\le N\le 65536$ (left panel) 
and the interaction case 
$U=2$, $U_r=1$ with system size $128\le N\le 2048$ (right panel). 
In the left panel also the classical decay based on the model of 
simple classical diffusion (with absorption for only the second 
particle) is shown.

The results shown in Fig.~\ref{fig12} are very similar to the results 
of Fig.~\ref{fig6} if one takes into account that all decay 
times are increased by a factor of 2, i.e. we have roughly 
$P_{\rm AS}(t)\approx P(t/2)$
where $P_{\rm AS}(t)$ is a curve of Fig.~\ref{fig12} for the 
asymmetric absorption of only the second particle and 
$P(t)$ is a curve of Fig.~\ref{fig6} for the absorption of both particles.

\begin{figure}
\begin{center}
\includegraphics[width=0.8\textwidth]{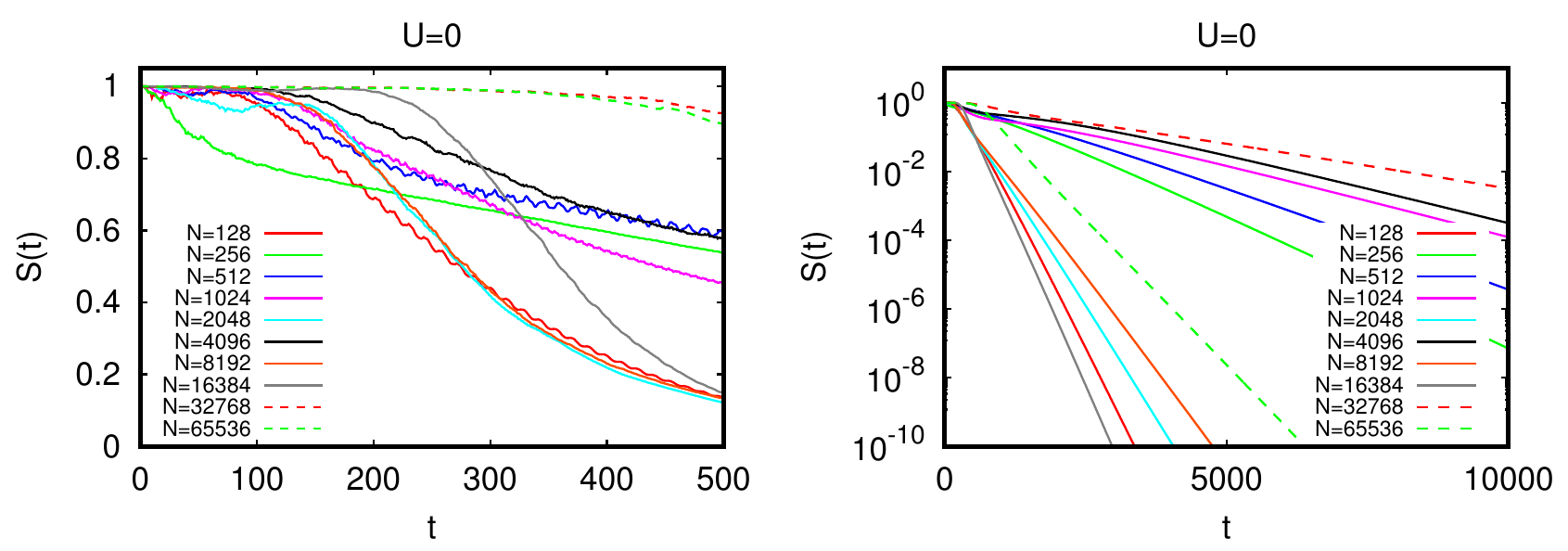}
\caption{
{\em Left:} Decay of the entropy of entanglement $S(t)$
for interaction value $U=0$, system size  $128\le N\le 65536$ 
and absorption only for the second particle. 
{\em Right:} As left panel but using a logarithmic scale for the entropy 
and an increased time range. 
}
\label{fig13}
\end{center}
\end{figure}

Fig.~\ref{fig13} shows for the case of asymmetric absorption, 
$U=0$ and different values of $N$ 
the decay of the entropy of entanglement $S(t)$ which converges also to 0 
for $t\to\infty$ for all cases. Qualitatively the overall (rather complicated) 
behavior is similar to the case of symmetric absorption shown 
in Fig.~\ref{fig7}, i.e. the decay times depend in a non-systematic 
way on $N$, sometimes with a late onset of the decay. However, 
cases for slow, fast or late decay happen for the same values of $N$ 
as in Fig.~\ref{fig7}. 

We find again that $\alpha_1(t)\to 1$ and $\alpha_2(t)\to 0$ for 
$t\to\infty$. The state $\ket{v_1(t)}$ is now of course different 
from $\ket{u_1(t)}$ but it is identical (up to a phase factor) 
for $t\to\infty$ to the state $\ket{v_1(t)}$ for the case of symmetric 
absorption while $\ket{u_1(t)}$ corresponds to the free (ergodic) 
quantum evolution of the first particle. (See below for the 
corresponding Husimi functions for some examples.) 

The efficient computation method for $U=0$ (see Appendix A.1) 
is also valid here. However, the triangular $2\times 2$ matrix $R_u$ 
associated to the first particle is now 
always the unity matrix since no orthogonalization procedure for the 
states $\ket{u_{1,2}}$ is necessary during the iteration 
(or in other words 
if it is done anyway one simply obtains for $R_u$ the unit matrix). 
However, the other matrix $R_v$ for the second particle has a non-trivial 
behavior and therefore the Schmidt decomposition still 
evolves in a non-trivial way similar to the scheme described in Appendix A.1. 

The state $\ket{u_1(t)}$ at arbitrary time $t$ 
is a linear combination of both initial states $\ket{u_{1,2}(t=0)}$ 
to which the free one-particle kicked rotator 
time evolution is applied but the coefficients of this linear combination 
depend on the way the second particle is absorbed (by the successive 
products of the matrix $R_v$). This situation is quite similar 
to the case of measurement of the second particle at 
some specific $p_2$ values studied in \cite{eprchaos} 
since such a measurement process can be viewed as an extreme case of 
absorption for all $p_2$ values being different from the measured value.

\begin{figure}
\begin{center}
\includegraphics[width=0.8\textwidth]{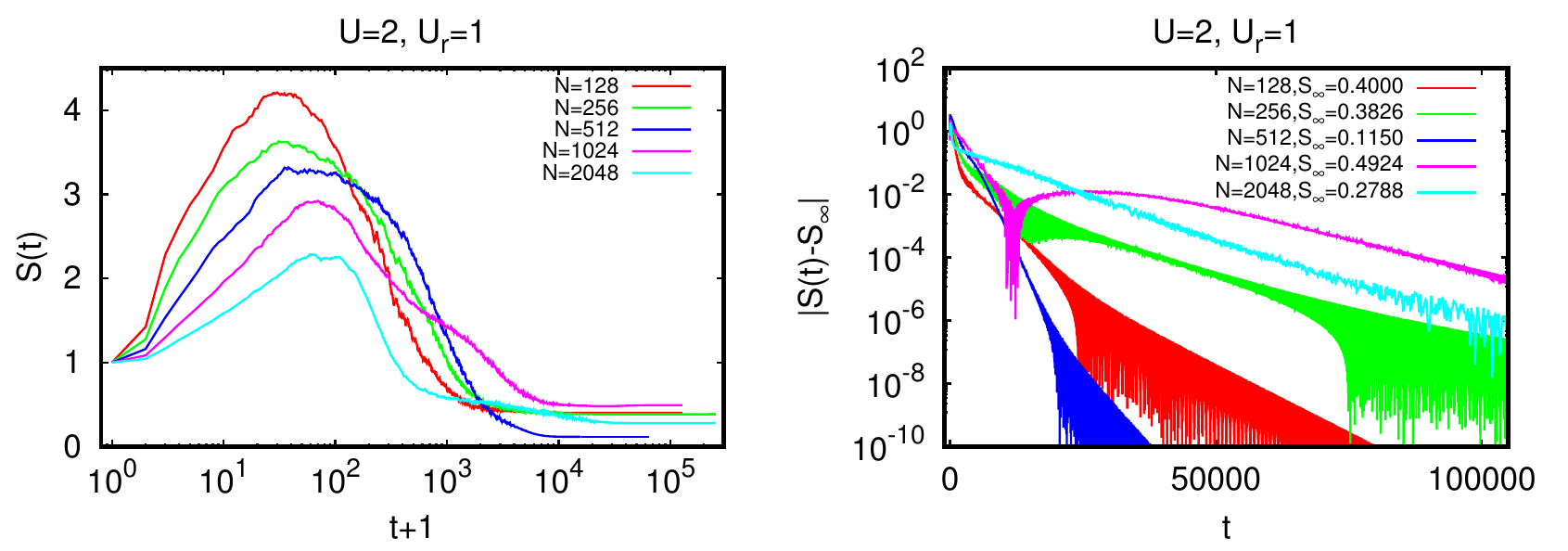}
\caption{
{\em Left:} Entropy of entanglement $S(t)$ versus iteration time 
for interaction values $U=2$, $U_r=1$, system size  $128\le N\le 2048$ 
and absorption only for the second particle. The horizontal axis is 
shown in logarithmic scale for $(t+1)$ (in order to keep the first 
data point at $t=0$ visible). 
{\em Right:} Decay of the difference $|S(t)-S_{\infty}|$ for the same values 
of interaction and system size with $S_{\infty}=\lim_{t\to\infty} S(t)$. 
The shown time scale is linear 
and a logarithmic scale for the vertical axis is used.
The modulus of the difference is taken because $S(t)$ is for certain 
time values $t$ below the limit value due to small 
amplitude oscillations for long time scales. 
The value of $S_\infty$ is significantly smaller than unity. 
}
\label{fig14}
\end{center}
\end{figure}

Fig.~\ref{fig14} shows for the asymmetric absorption case 
the behavior of the entropy $S(t)$ for the 
case of $U=2$, $U_r=1$ and system size $128\le N\le 2048$.
Initially the entropy increases to maximum values between 
$S_{\rm max}\approx 4.2$ at $t\approx 25$ (for $N=128$) and 
$S_{\rm max}\approx 2.3$ at $t\approx 60$ (for $N=2048$) 
and then it decays for $t\to\infty$ exponentially
to a limit value $S_{\infty}$ which is now significantly smaller than 
unity (but larger than zero). The values of $S_{\infty}$ for 
each case of $N$ are also given in the right panel 
of Fig.~\ref{fig14} which shows the (modulus of the) 
difference $S(t)-S_{\infty}$ 
versus time on a longer time scale and logarithmic scale for the 
vertical axis. As in the case of symmetric absorption one 
also observes that for long time scales there are 
small amplitude oscillations of $S(t)$ around $S_{\infty}$.
However, now the convergence of the entropy is considerably slower
than for the symmetric absorption case shown in Fig.~\ref{fig8}, 
especially for $N=1024$ and $N=2048$.

\begin{figure}
\begin{center}
\includegraphics[width=0.8\textwidth]{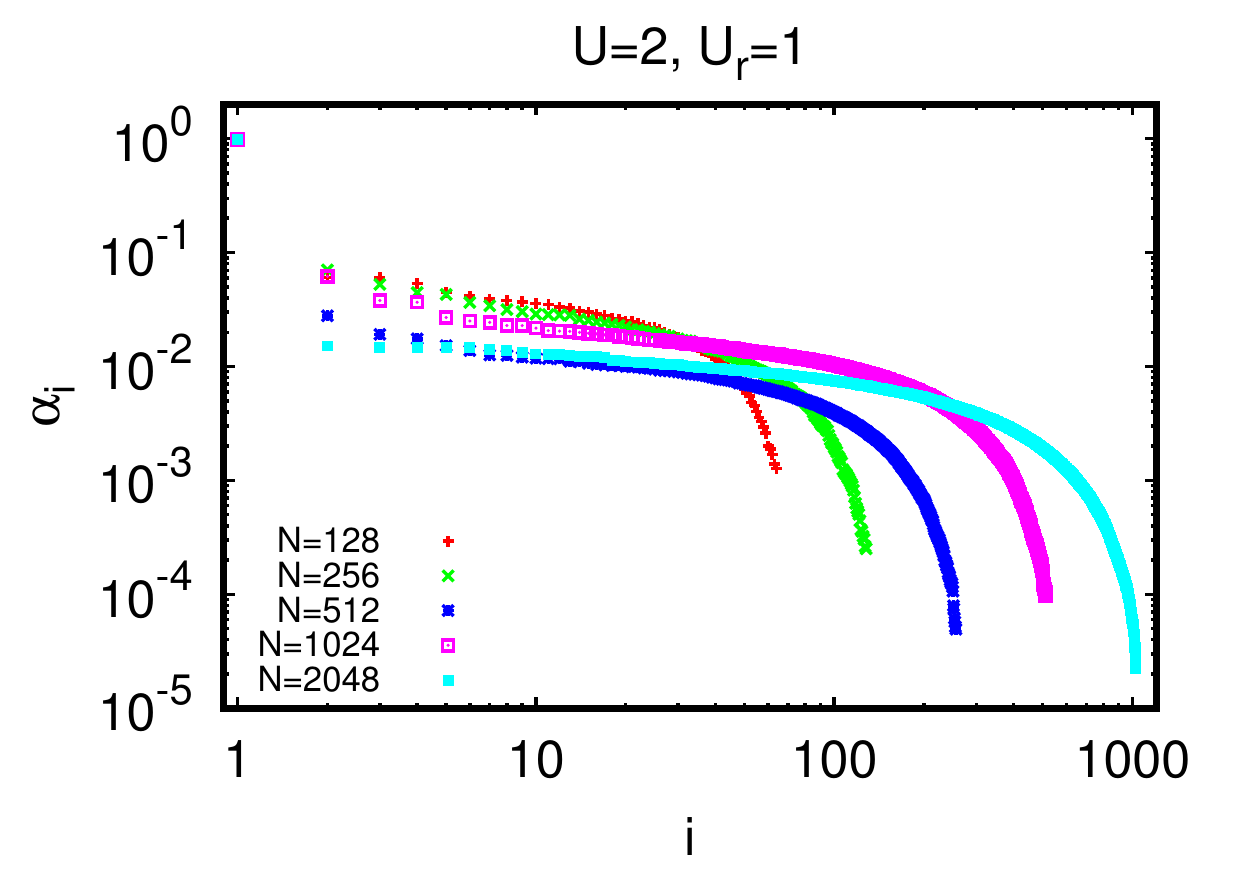}
\caption{
Singular values $\alpha_i$ (appearing in the  Schmidt decomposition of 
the limit state $\lim_{t\to\infty}|\psi(t)\rangle$) versus index $i$
for interaction values $U=2$, $U_r=1$, system size $128\le N\le 2048$ 
and absorption only for the second particle. 
Both axis are shown on a logarithmic scale. The values of the top 
singular value are 
$\alpha_1=0.98186,\,0.98337,\,0.99612,\,0.98136,\,0.99107$
for $N=128,\,256,\,512,\,1024,\,2048$ respectively.
}
\label{fig15}
\end{center}
\end{figure}

For $U=2$, $U_r$ and the asymmetric absorption, also 
all singular values contribute to the limit state 
as can be seen in Fig.~\ref{fig15}. However, 
now, due to the obvious absence of antisymmetry in the limit state (with 
respect to particle exchange) the singular values are no longer 
pairwise degenerate and the top singular 
value $\alpha_1\approx 0.98$--$0.99$ (see caption of Fig.~\ref{fig15}
for more precise values) is dominant while 
$\alpha_i< 0.062$ for $i\ge 2$ (and $N=1024$). However, despite 
the dominating first singular value the resulting entropy 
(of the limit state) is still considerably larger 
than zero with values $\approx0.3-0.5$ (except for $N=512$ where 
the limit value is $\approx 0.1$).

\begin{figure}
\begin{center}
\includegraphics[width=0.8\textwidth]{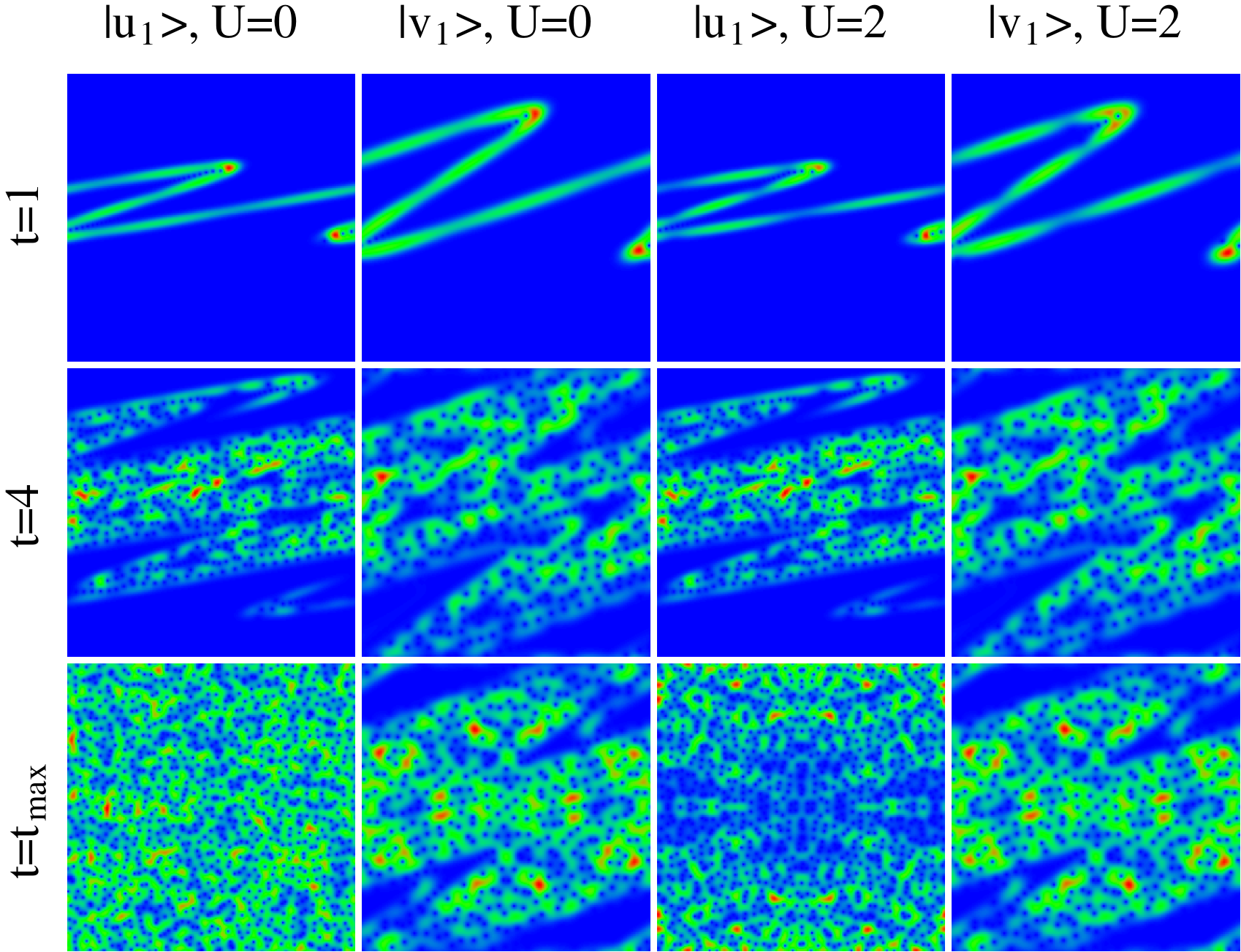}
\caption{
Husimi functions of the Schmidt states $|u_1\rangle$, $|v_1\rangle$, 
for $U=0$ or $U=2$ (with $U_r=1$), $N=1024$ 
and absorption only for the second particle 
at three iteration times $t=1$, $t=4$ and 
$t=t_{\rm max}$ with $t_{\rm max}=16384$ for $U=0$ or $t_{\rm max}=131072$ 
for $U=2$. 
The horizontal axis corresponds to the phase  $\theta\in[-\pi,\pi]$ and 
the vertical axis to the momentum $p_{\rm cl}\in[-2p_{\rm max},2p_{\rm max}]$ 
(for $|u_1\rangle$) or $p_{\rm cl}\in[-p_{\rm max},p_{\rm max}]$ 
(for $|v_1\rangle$) where $p_{\rm max}=(2\pi)\cdot 2.228$ 
is the momentum absorption border in classical units (for the second 
particle) and $2p_{\rm max}$ is the maximal 
classical momentum value for the full phase
space in absence of absorption (for the first particle). 
The colors red/green/blue correspond to 
maximal/medium/minimal values of the Husimi function. 
}
\label{fig16}
\end{center}
\end{figure}

As in the last section, 
we have also computed for $U=0$ and the interaction cases the first 
Schmidt vectors $\ket{u_{1,2}}$, $\ket{v_{1,2}}$ and the associated 
Husimi functions \cite{qcfrahm,husimi,eprchaos}. 
The vectors $\ket{u_1(t)}$, $\ket{v_1(t)}$ for $U=0$ and 
$U=2$, $U_r=1$ (all for $N=1024$) 
are shown in Fig.~\ref{fig16} for three time values 
being $t=1$, $t=4$ and some very long time being $t=t_{\rm max}=16384$ 
for $U=0$ or $t=t_{\rm max}=131072$ for $U=2$, $U_r=1$. 
Again for $t=1$, these states occupy the same manifold but for the two 
interaction cases the densities at some positions on 
this manifold are reduced in comparison to the non-interacting case.
(Note that the available phase space for $\ket{v_1(t)}$ is reduced by a factor 
of 2 concerning the maximal $p$ value as compared to $\ket{u_1(t)}$.)

Again for $t=4$ the phase space structure is quite complicated but rather 
similar between the interacting and non-interacting cases for both 
states $\ket{u_1}$ and $\ket{v_1}$ respectively. 

At long times the state $\ket{u_1}$ of $U=0$ seems to be ergodic 
while the state $\ket{v_1}$ coincides (quite exactly and for both 
$U=0$ and $U=2$, $U_r=1$) with the 
state $\ket{v_1}$ for the symmetric absorption case ($U=0$) visible 
in Fig.~\ref{fig10} (see $\ket{u_1}$ in this figure which 
coincides with $\ket{v_1}$ for the symmetric absorption case at $U=0$). 
However, for $U=2$, $U_r=1$ the state $\ket{u_1}$ (at long times) 
is quite strange with a strongly enhanced probability in the absorption 
area $|p|>p_{\rm max}$ of the other particle. 
Furthermore, also in the region $|p|<p_{\rm max}$ the first particle 
has larger (smaller) probability values at classical positions 
where the second particle is absent (present). 
This indicates that the first particle is somehow 
repelled by the interaction from the second particle 
which cannot enter the absorption area and which is in a quite stable limit state 
$\ket{v_1}$.

\begin{figure}
\begin{center}
\includegraphics[width=0.8\textwidth]{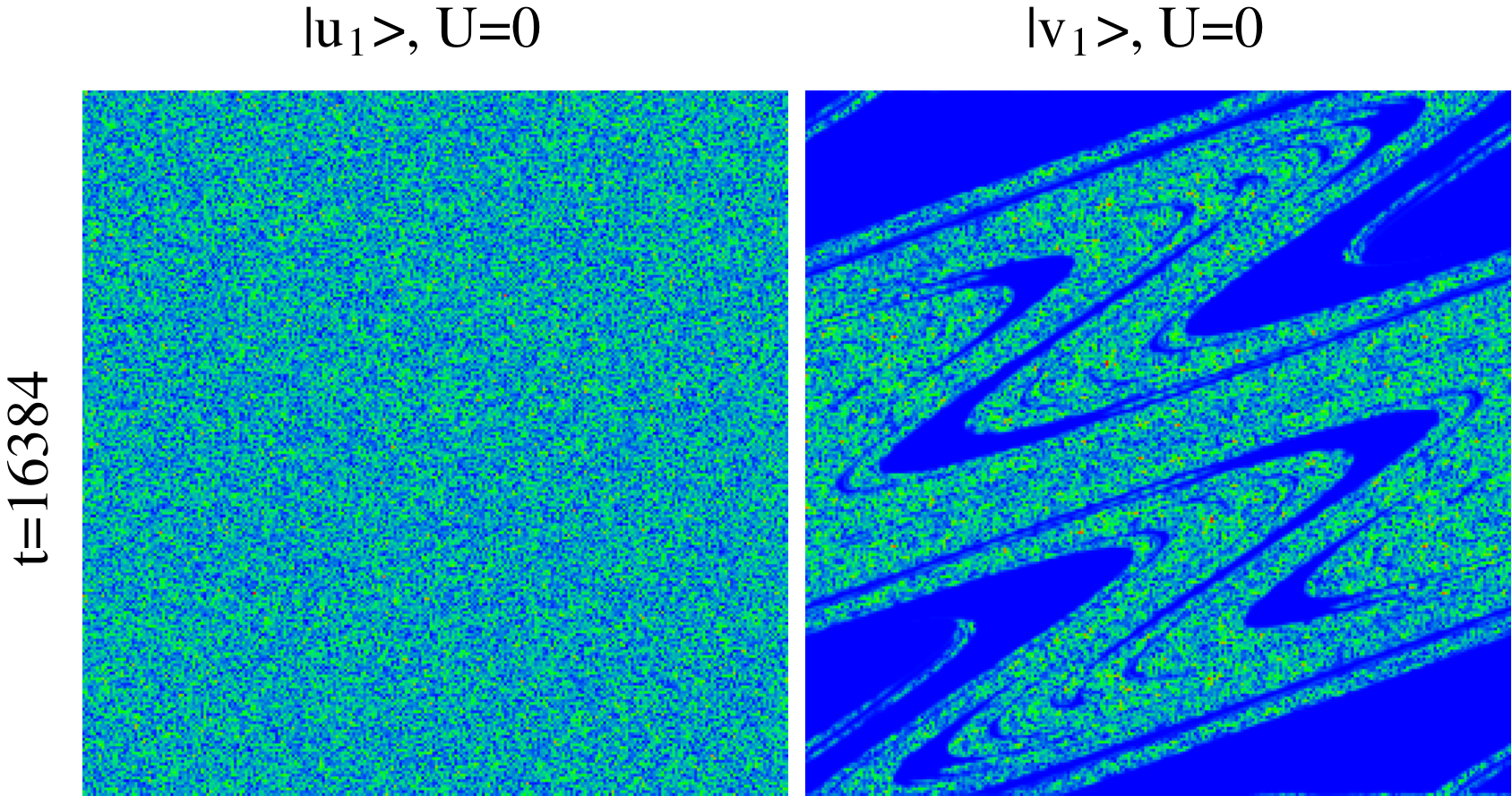}
\caption{
Husimi functions of the Schmidt states $|u_1\rangle$, $|v_1\rangle$, 
for $U=0$, $N=65536$ 
and absorption only for the second particle 
at final iteration time $t=16384$. The signification 
of both axes and the color codes is the same as in Fig.~\ref{fig16}.
}
\label{fig17}
\end{center}
\end{figure}

In Fig.~\ref{fig17}, we also show for the asymmetric absorption case 
the Husimi functions of $\ket{u_1}$ and $\ket{v_1}$
for $U=0$ and a larger system size $N=65536$ at final iteration time $t=16384$.
Similarly to the case $N=1024$ the state $\ket{u_1}$ is ergodic and 
$\ket{v_1}$ coincides (quite precisely) with the corresponding state 
for the symmetric absorption case (see state $\ket{u_1}$ of 
Fig.~\ref{fig11}).   The Husimi function of
the second absorbed particle has a fractal structure
corresponding to a fractal repeller of non-escaping classical orbits,
while the Husimi function of the first particle is homogeneous in the phase space.

\section{Poincar\'e recurrences of entanglement with absorption of one particle 
for $K=2.5$}

In this Section, we consider the case of asymmetric absorption only for the 
second particle for the classical chaos parameter $K=2.5$ and 
same value of $k=L/4=N/8$ implying that now 
$p_{\rm max}=5=(2\pi)\cdot 0.7958$ corresponds to $0.7958$ classical 
momentum cells between $0$ and $p_{\rm max}$. 
The statistics of Poincar\'e recurrences for this case was studied 
in \cite{maspero2} and here the phase space is mixed with remarkable 
stable islands (see Husimi figures below). 
This leads to a power law decay of $P(t)\propto t^{-1}$ 
due to resonant classical modes outside but very close to the stable 
islands \cite{maspero2}. Similarly as in \cite{maspero2}, we choose 
as initial Schmidt states localized states in momentum at (quantum) 
values close to $L/3$, more precisely~:
$\ket{u_1(0)}=\ket{p_1=20\Delta p}$, $\ket{u_2(0)}=\ket{p_1=21\Delta p}$, 
$\ket{v_1(0)}=\ket{p_2=21\Delta p}$ and $\ket{v_2(0)}=\ket{p_2=22\Delta p}$ 
with $\Delta p=N/128=L/64$ being the same scaling factor used 
in (\ref{equ1})-(\ref{eqv2}). For these values of $p_{1,2}$ the initial 
momentum lines have no intersections with the stable islands. 
Now, the entropy $S(t)$ does not converge (for $U=0$) or not very 
well (for $U=2$) and we choose for all cases a maximal iteration time 
of $t_{\rm max}=2^{18}$.

\begin{figure}
\begin{center}
\includegraphics[width=0.8\textwidth]{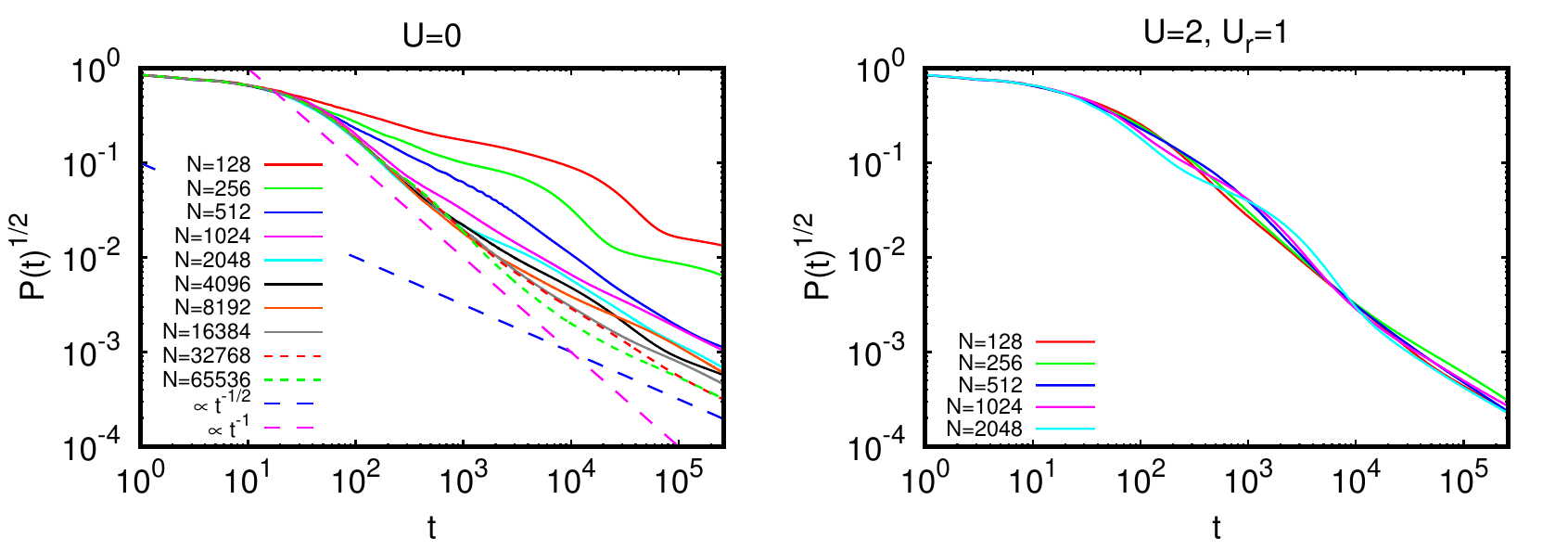}
\caption{
{\em Left:} Decay of $P(t)^{1/2}=\||\psi(t)\rangle\|$ where the norm 
is obtained before renormalization of the state  $|\psi(t)\rangle$
for interaction value $U=0$, system size  $128\le N\le 65536$, $K=2.5$ 
and absorption only for the second particle. 
The long dashed blue (pink) straight line corresponds to the 
power law $\propto t^{-1/2}$ ($\propto t^{-1}$) 
for comparison. 
{\em Right:} Decay of $P(t)^{1/2}=\||\psi(t)\rangle\|$ 
for interaction values $U=2$, $U_r=1$ and system size $128\le N\le 2048$.
Both panels are in a double logarithmic representation. 
}
\label{fig18}
\end{center}
\end{figure}

Fig.~\ref{fig18} shows for this case 
the quantity $P(t)^{1/2}$ obtained from the 
quantum iteration (\ref{eqtimeevolut}) for 
$U=0$ with system size $128\le N\le 65536$ (left panel) 
and the interaction case 
$U=2$, $U_r=1$ with system size $128\le N\le 2048$ (right panel). 
Now a double logarithmic representation is chosen since the decay 
is for both cases close to the power law $P(t)\propto t^{-1}$ 
($P(t)\propto t^{-1/2}$) for intermediate (longer) time scales confirming 
the findings of \cite{maspero2} for $U=0$. For the interaction case $U=2$ 
the decay curves for modest values of $N$ between $128$ and $2048$ 
do not strongly depend on $N$ (in contrast to $U=0$ for these size values) 
and are actually closer to the curves of large $N$ values 
for $U=0$ and both types of power law decay (depending on longer
or intermediate time scales). 

\begin{figure}
\begin{center}
\includegraphics[width=0.8\textwidth]{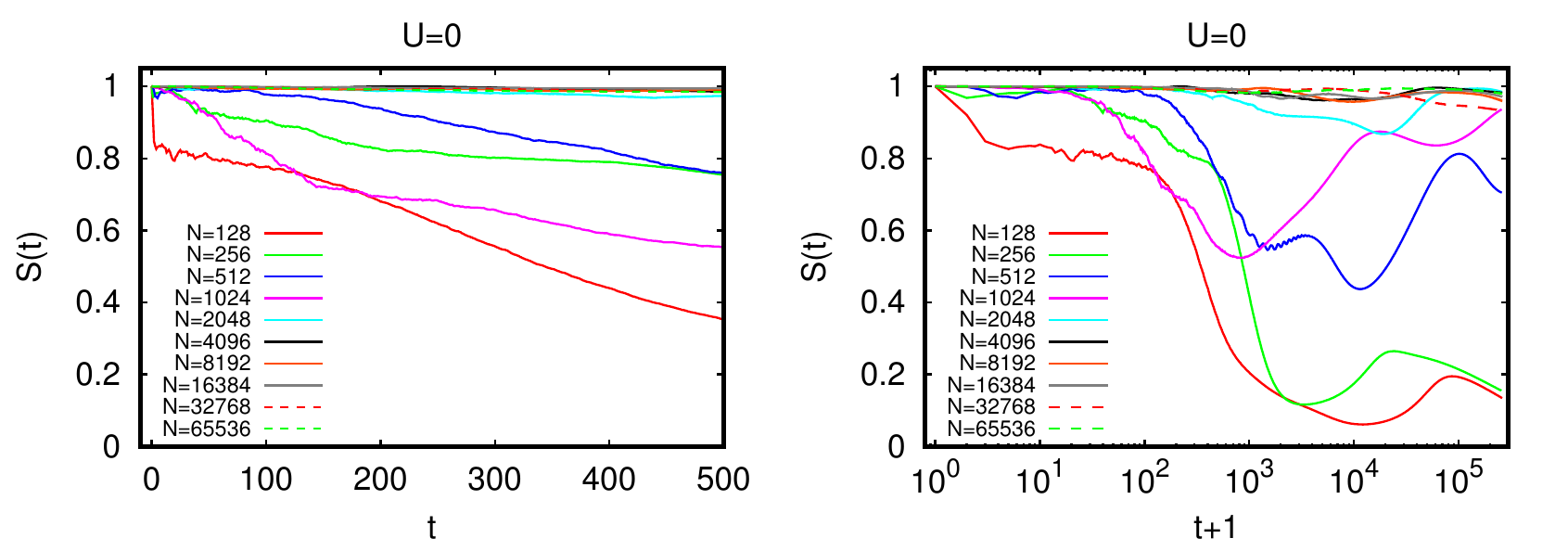}
\caption{
{\em Left:} Time dependence of the entropy of entanglement $S(t)$
for interaction value $U=0$, system size  $128\le N\le 65536$, $K=2.5$  
and absorption only for the second particle. 
{\em Right:} As left panel but using a logarithmic scale 
on the horizontal axis in $(t+1)$ (in order to keep the first 
data point at $t=0$ visible). 
}
\label{fig19}
\end{center}
\end{figure}

Fig.~\ref{fig19} shows for asymmetric absorption, $K=2.5$, 
$U=0$ and different values of $N$ the behavior of the entropy of 
entanglement $S(t)$. The entropy seems to decay 
for short time scales and smaller $N$ values but has a more complicated 
behavior for larger time scales where the entropy may re-increase even up 
the initial unity value for some cases. For larger $N$ values the 
initial decay is very slow and small in amplitude. 

It seems that due to the weak decay of $P(t)$ the Schmidt states 
$\ket{v_{1,2}}$ of the second particle retain partly their orthogonality 
such that both singular values stay close to their initial 
values $\alpha_1(t)\approx \alpha_2(t)\approx 1/\sqrt{2}$ implying 
that the entanglement is rather well conserved. However, one might 
expect that at extremely long exponential time scales (in $N$) 
the entropy should also decay to $0$ in a similar way the quantum decay 
of $P(t)$ should become exponential for such time scales \cite{maspero2}. 

The efficient computation method for $U=0$ (see Appendix A.1) 
was again used for the data of Fig.~\ref{fig19}. 
In a similar way, as for the case $K=7$ 
(with absorption only for the second particle), 
the state $\ket{u_1(t)}$ at arbitrary time $t$ 
is a linear combination of both initial states $\ket{u_{1,2}(t=0)}$ 
to which the free one-particle kicked rotator 
time evolution is applied but the coefficients of this linear combination 
depend on the way the second particle is absorbed. 
The same holds for $\ket{v_{1,2}(t)}$ which are also given 
by a linear combination of both initial states $\ket{v_{1,2}(t=0)}$ 
to which the free one-particle kicked rotator {\em with} absorption 
is applied. However, while for $K=7$ 
both vectors converged quite well to the first two modes of 
the non-unitary (one-particle) iteration operator, this is not the 
case for $K=2.5$ due to the exponentially long time scale 
for this type of convergence. 

\begin{figure}
\begin{center}
\includegraphics[width=0.8\textwidth]{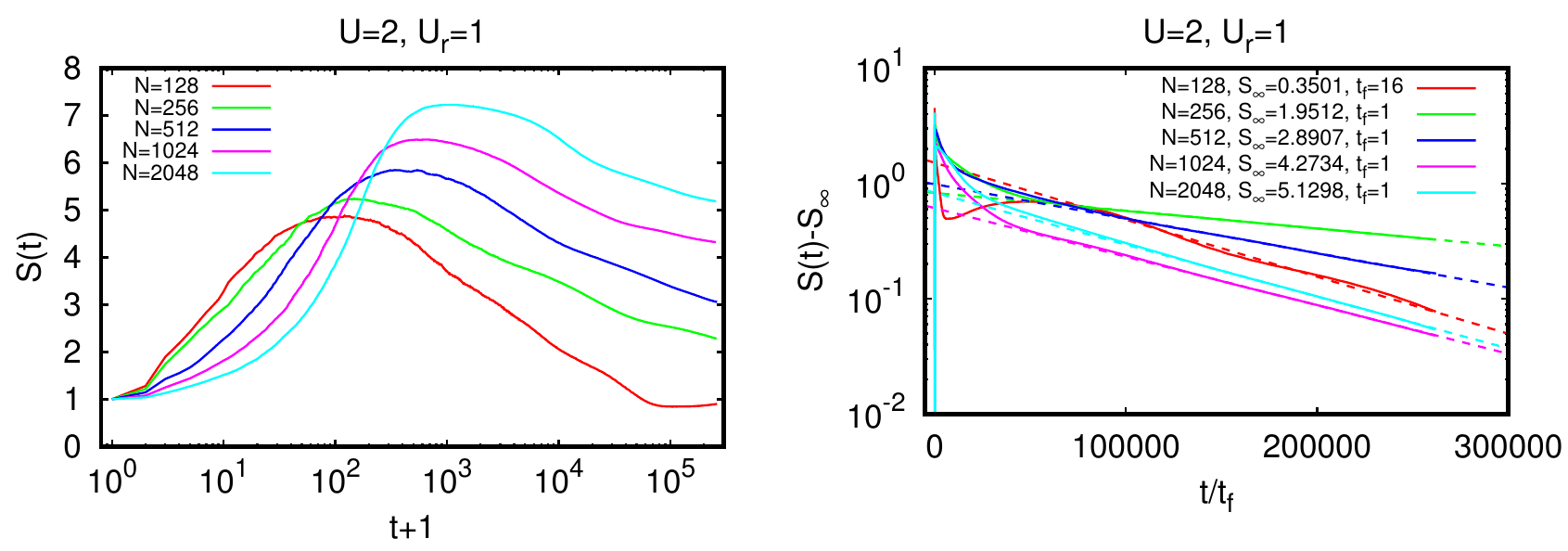}
\caption{
{\em Left:} Entropy of entanglement $S(t)$ versus iteration time 
for interaction values $U=2$, $U_r=1$, system size  $128\le N\le 2048$, 
$K=2.5$  
and absorption only for the second particle. The horizontal axis is 
shown in logarithmic scale for $(t+1)$ (in order to keep the first 
data point at $t=0$ visible). 
{\em Right:} Decay of the difference $S(t)-S_{\infty}$ for the same values 
of interaction and system size where $S_{\infty}$ 
is determined from the exponential fit 
with a constant term 
$S(t)=S_\infty+S_A\,\exp(-t/t_S)$ for the time interval 
$2^{17}\le t\le 2^{18}$ ($2^{20}\le t\le 2^{22}$) 
for $N\ge 256$ ($N=128$). 
The resulting fit values are: 
$S_\infty=0.3501\pm0.0039$, $S_A=1.515\pm0.010$, 
$t_S=(1.403 \pm 0.017)\times 10^{6}$ for $N=128$; 
$S_\infty=1.9512\pm0.0009$, $S_A=0.8298\pm0.0006$, 
$t_S=(2.808 \pm 0.006)\times 10^{5}$ for $N=256$; 
$S_\infty=2.8907\pm0.0026$, $S_A=0.9865\pm 0.0029$, 
$t_S=(1.452 \pm 0.014)\times 10^{5}$ for $N=512$; 
$S_\infty=4.2734\pm0.0004$, $S_A=0.6117\pm 0.0018$, 
$t_S=(1.030 \pm 0.004)\times 10^{5}$ for $N=1024$;
$S_\infty=5.1298\pm0.0004$, $S_A=0.8328\pm 0.0025$, 
$t_S=(9.63\pm 0.03)\times 10^{4}$ for $N=2048$. 
The straight dashed lines of same color show for 
each value of $N$ the resulting fit curve. 
The inverse scaling factor $t_f$ on the horizontal time axis 
is 1 (16) for $N\ge 256$ ($N=128$) 
corresponding to a maximal iteration time $t_{\rm max}=2^{18}$ 
($t_{\rm max}=2^{22}$). 
The shown time scale is linear and a 
logarithmic scale for the vertical axis is used.
}
\label{fig20}
\end{center}
\end{figure}

Fig.~\ref{fig20} shows for the asymmetric absorption case 
and $K=2.5$ the behavior of the entropy $S(t)$ for 
$U=2$, $U_r=1$ and system size $128\le N\le 2048$.
Initially, the entropy increases to maximum values between 
$S_{\rm max}\approx 5$ at $t\approx 10^2$ (for $N=128$) and 
$S_{\rm max}\approx 7.2$ at $t\approx 10^3$ (for $N=2048$). 
Interestingly, now 
the intermediate (first) maximum values of $S(t)$ are larger for larger values 
of $N$ (instead of smaller values of $N$ as for $K=7$). 
For $t\to\infty$ the entropy decays 
exponentially to a limit value $S_{\infty}$ which 
is now significantly larger than unity (except for $N=128$) 
but the decay times are very long of the order of $t_S\sim 10^5$ or even 
$t_S\sim 10^6$ for $N=128$ and the limit value 
is obtained by an exponential fit with a constant term for the 
time interval $2^{17}\le t\le 2^{18}$ ($2^{20}\le t\le 2^{22}$ for $N=128$). 
We note that for $N=128$ the entropy has a minimum at $t\approx 10^5$ 
and a second maximum at $t\approx 10^6\approx 2^{20}$ and some slight 
oscillatory behavior around the fit line is visible. Due to the late 
onset of the decay of $S(t)$ for $N=128$, we have increased the 
maximal iteration time here to $t_{\rm max}=2^{22}$ and in the (right panel) 
of Fig.~\ref{fig20} the time axis for $N=128$ is rescaled by a factor 
of $1/16$. 

In principle, due to the obtained behavior at $N=128$, we cannot exclude for certain that 
also for $N\ge 256$ the entropy might re-increase at some long time with 
a second maximum. However, the exponential fits shown in the figure are of 
good quality and for all time values $t>10^2$ 
the values of $S(t)$ are clearly above the given value of $S_\infty$. 

For the case of this Section
we present additional results for the spectrum of singular values $\alpha_i$,
and Husimi functions in Figs.~S2, S3, S4 of SupMat.

\section{Discussion}

In this work we studied the properties of
entanglement of two particles
using such tools of quantum chaos as Loschmidt echo
and Poincar\'e recurrences with absorption.

We find that the Loschmidt echo of entanglement
$G(t_r) = \alpha t_r$ is characterized
by a linear growth at small times $t_r$
with the rate $\alpha$ being proportional to the square of the perturbation
as for the Fermi golden rule ($\alpha \propto (\Delta U)^2$).
We attribute such a dependence to the perturbation
corrections to the Kolmogorov-Sinai entropy
which is at the origin of exponentially fast
wave packet spreading.

For the case of Poincar\'e recurrences of entanglement, 
we find different unusual regimes for the decay of the entropy of
entanglement $S(t)$. For the case of absorption of both particles 
and in absence of interactions 
$S$ always decays to zero. 
However this decay
is very slow and depends on the system size in a complex manner. 
Furthermore, it is related to the gap between the first two decay 
rates obtained from the first two complex eigenvalues of the 
non-unitary one-particle iteration operator. 
In presence of interactions $S$ decays to a finite value
and the decay rate to it can also be very small
depending on the singular values of Schmidt decomposition.
The Husimi function of the asymptotic state
has a fractal repeller structure for both particles.
Furthermore, the two particle state converges to an anti-symmetric 
state with respect to particle exchange providing double degenerate 
singular values appearing in the Schmidt decomposition. The 
entropy $S_\infty$ of this limit state is either significantly 
larger than or equal to unity depending on the interaction range 
$U_r=1$ or $U_r=0$ respectively. 

For the case of absorption of only one particle
the interactions significantly modify the asymptotic
Husimi function of the particle without absorption, e.g., 
effective repulsion structure from the other absorbed particle 
or faster penetration inside stable islands. 
The entropy of entanglement of two interacting particles
also decays to a finite value. 
Depending on the chaos parameter the limit entropy $S_\infty$ 
may be significantly below unity ($K=7$) or quite large 
up to $S\approx 4-5$ ($K=2.5$ and largest system sizes). 
For the case $K=2.5$ the classical phase space has
a hierarchical structure of integrable islands and 
here the time scale for the exponential entropy decay 
is enormously large.

Our studies are done for the case of the
quantum Chirikov standard map which describes a generic behavior
in the regime of quantum chaos. Indeed, other systems, as e.g.
the kicked top, actively studied by Fritz Haake and his colleagues
(see e.g. \cite{kickedtop} and Refs. therein), can be locally described by
the Chirikov standard map \cite{haakedls}.
Due to this reason we expect that the obtained results
are generic.

\ack
We enjoyed multiple friendly and chaotic discussions with Fritz Haake
starting for an international workshop on quantum optics
in Ustron, Poland in the fall of 1985
at the rise of perestroika.

This research was supported in part through the grant 
NANOX $N^o$ ANR-17-EURE-0009, (project MTDINA) in 
the frame of the Programme des Investissements d’Avenir, France;
the work is also supported in part by the ANR FRANCE OCTAVES (ANR-21-CE47-0007).
This work was granted access to the HPC resources of 
CALMIP (Toulouse) under the allocation 2021-P0110.

\section*{Appendix A: Schmidt decomposition and singular value decomposition}
\setcounter{equation}{0}
\renewcommand{\theequation}{A\arabic{equation}}
\setcounter{figure}{0}
\renewcommand\thefigure{A\arabic{figure}}
\renewcommand{\figurename}{Appendix Figure}

\subsection*{A.1: General case}

The singular values $\alpha_i(t)$ and Schmidt vectors $\ket{u_i(t)}$ 
and $\ket{v_i(t)}$ in (\ref{eqschmidtgeneral}) 
can be computed from the singular value decomposition 
of the ``matrix'' $\psi(p_1,p_2)$ with $p_1$, $p_2$ being the matrix indices. 
For this we introduce the density matrix of the first (second) particle 
$\rho_1(p_1,q_1)=\sum_{p_2} \psi(p_1,p_2)\psi(q_1,p_2)^*$
[or $\rho_2(p_2,q_2)=\sum_{p_1} \psi(p_1,p_2)\psi(p_1,q_2)^*$] where the 
$p_2$-- ($p_1$--) sum corresponds to the partial trace over the second 
(first) particle. 

In matrix notation, we can also write $\rho_1=\psi\psi^\dagger$
(or $\rho_2=\psi^\dagger\psi$). Note that for the asymmetric case 
when only one particle is absorbed the matrix $\psi$ may be rectangular 
and in this case $\rho_1$ and $\rho_2$ may be of different size.
Both matrices $\rho_1$ and $\rho_2$ are Hermitian and have real eigenvalues. 
Let $u_i$ be a normalized eigenvector of $\rho_1$ with such an 
eigenvalue $\lambda_i$. 
Since \begin{equation}
\label{eqlambdai}
\lambda_i=\langle u_i\ket{\rho_1 u_i}=
\langle u_i\ket{\psi\psi^\dagger u_i}=\langle\psi^\dagger u_i
\ket{\psi^\dagger u_i}=\|\psi^\dagger u_i\|^2\ge 0 
\end{equation}
we write 
$\lambda_i=\alpha_i^2$ with real $\alpha_i\ge 0$. 
For $\alpha_i>0$ let us define the (normalized) 
vector $v_i=\psi^\dagger u_i/\alpha_i$ 
such that:
\begin{equation}
\label{eqrho2vec}
\rho_2 v_i=(\psi^\dagger\psi)\psi^\dagger u_i/\alpha_i
=\psi^\dagger(\psi\psi^\dagger) u_i/\alpha_i
=\psi^\dagger(\alpha_i^2) u_i/\alpha_i
=\alpha_i^2 v_i
\end{equation}
showing that $v_i$ is an eigenvector of $\rho_2$ with the same 
eigenvalue $\alpha_i$. Inversely, for a given eigenvector $v_i$ of $\rho_2$ 
with eigenvalue $\alpha_i^2>0$ 
one can obtain by $\tilde u_i=\psi v_i/\alpha_i$ a normalized 
eigenvector of $\rho_1$ (which may be different from $u_i$ 
for the case of a degenerate eigenvalue). 
Therefore, $\rho_1$ and $\rho_2$ have the same 
non-vanishing eigenvalues. Both of them may also have the eigenvalue zero 
but for the case of different matrix sizes it is possible that 
only one of them has the eigenvalue zero while the other does not. 
Using the eigenvectors $u_i$ and $v_i$, we construct unitary matrices 
$u$ and $v$ containing these vectors in their columns respectively. 
(For the case of some $\alpha_i=0$, we add additional column vectors 
orthogonal to the eigenvectors for $\alpha_i>0$ to 
obtain full squared unitary matrices.)
This provides the ``singular value decomposition'' 
$\psi=u\alpha\, v^\dagger$ where $\alpha$ is a matrix with diagonal 
elements $\alpha_i$ and zero non-diagonal elements (this matrix 
may be rectangular of the same size as $\psi$).

This scheme provides also a numerical procedure to compute the singular value 
decomposition, by first diagonalizing $\rho_1$ which gives the unitary 
matrix $u$ with columns vectors $u_i$ being the eigenvectors of $\rho_1$. 
The singular values $\alpha_i$ are obtained as the 
norm $\alpha_i=\|\psi^\dagger u_i\|$ (which is numerically more precise 
for small $\alpha_i$ than taking $\alpha_i=\sqrt{\lambda_i}$ if 
$\lambda_i$ is the numerically obtained eigenvalue of $\rho_1$).
The eigenvectors $v_i$ of $\rho_2$ are then constructed as above from $u_i$, 
with eventually adding further orthogonal vectors for the case 
$\alpha_i=0$. 
Once the singular value decomposition is known one chooses the columns 
of $u$ for $\ket{u_i}$ and the rows of $v^\dagger$ (i.e. columns of $v^*$) 
for $\ket{v_i}$ to obtain the Schmidt decomposition (\ref{eqschmidtgeneral}). 

For the case of a square (complex) matrix $\psi$ 
(absorption of both particles) being skew-symmetric $\psi^T=-\psi$ 
(e.g. the anti-symmetric state with respect 
to particle exchange for $t\to\infty$ and $U\neq 0$), 
there is a mathematical theorem \cite{rogerhorn} stating that the 
non-vanishing singular values come in degenerate pairs, i.e. the associated 
dimensions of the eigenspaces of $\rho_{1,2}$ are even. For the simple case of 
a {\em real} skew-symmetric matrix this 
is quite obvious since the matrix $\psi$ is
then also anti-Hermitian ($\psi^\dagger=-\psi$) with purely imaginary 
eigenvalues $\pm i\alpha_j$ which also come in complex conjugated pairs 
(since $\psi$ is real). The matrix relation that diagonalizes 
$\psi=uDu^\dagger$ (with $D$ being the diagonal matrix with entries 
$\pm i\alpha_j$) becomes immediately the singular value 
decomposition $\psi=u\alpha v^\dagger$ 
with $v^\dagger=Iu^\dagger$ where $I$ is a diagonal 
matrix with entries $\pm i$. 

However, the degeneracy for $\alpha_i>0$ is also valid for a 
{\em complex} skew-symmetric matrix $\psi$ but in this case there is no simple 
link between the complex eigenvalues of $\psi$ (which actually 
come in pairs of opposite sign) and its singular values. To get a 
simple understanding of this 
theorem (shown in \cite{rogerhorn}), we note that due to $\psi^T=-\psi$
(now for the complex case) we have: $\rho_2=\psi^\dagger\psi
=(-\psi)^*(-\psi^T)=(\psi\psi^\dagger)^*=\rho_1^*$. 
Therefore, if $u_i$ is an eigenvector of $\rho_1$ 
we know that $u_i^*$ is an eigenvector of $\rho_2$ with the same eigenvalue 
and for $\alpha_i>0$ we can construct, according to the argument 
presented above, by $\tilde u_i=\psi u_i^*/\alpha_i$ an eigenvector of 
$\rho_1$. It turns out that this new eigenvector cannot be a multiple 
of the initial eigenvector $u_i$. Assuming $u_i=C\tilde u_i$, with some phase 
factor $C$, we obtain:
\begin{equation}
\label{eqcontradiction}
u_i=C\psi u_i^*/\alpha_i=|C|^2\psi\psi^* u_i/\alpha_i^2=
|C|^2 (-\psi\psi^\dagger) u_i/\alpha_i^2=-u_i
\end{equation}
which is impossible. Furthermore, $\tilde u_i$ is even orthogonal to $u_i$ 
since
\begin{eqnarray}
\label{eqtildeorth}
\langle \tilde u_i\ket{u_i}&=&\langle \psi u_i^*/\alpha_i\ket{u_i}
=\langle u_i^*\ket{\psi^\dagger u_i/\alpha_i}\\
\nonumber
&=&\langle u_i^*\ket{-\psi^* u_i/\alpha_i}
=-\langle u_i^*\ket{\tilde u_i^*}=-\langle \tilde u_i\ket{u_i}
\end{eqnarray}
implying $\langle \tilde u_i\ket{u_i}=0$. Therefore, for a given 
eigenvector $u_i$ we can construct a second linearly independent 
eigenvector $\tilde u_i$. Applying the same construction scheme 
to $\tilde u_i$ we find $-u_i$ 
with a similar calculation as in 
(\ref{eqcontradiction}). Therefore the singular values different 
from zero are at least double degenerate and it is not difficult 
to argue that higher degeneracies must be even.
Fig.~\ref{fig9} provides a numerical illustration of these degeneracies for 
the limit state of one of the interaction cases. 

\subsection*{A.2: Recomputation of Schmidt decomposition for $U=0$}

In absence of interaction the quantum iteration with absorption 
(\ref{eqtimeevolut}) does not increase the number of Schmidt 
components in the initial condition, i.e., in (\ref{eqschmidtgeneral}) 
there are only only 
two non-vanishing singular values being $\alpha_1$ and $\alpha_2$ and 
$\alpha_i=0$ for $i=3,\ldots,L$.

To see this, let us assume that we know the Schmidt decomposition 
(\ref{eqschmidtgeneral}) with $L=2$ components of a state 
$\ket{\psi(t)}$ at a given iteration time $t$. In absence 
of interaction the iteration operator (with absorption) acts independently 
on the Schmidt states $\ket{u_{1,2}}\to \ket{\bar u_{1,2}}$
and $\ket{v_{1,2}}\to \ket{\bar v_{1,2}}$. The new states 
are neither normalized neither orthogonal due to the non-unitarity of 
the iteration operator. However, one can normalize 
$\ket{\bar u_1}\to\ket{\hat u_1}$ and $\ket{\bar v_1}\to\ket{\hat v_1}$ 
and orthogonalize (by the usual Gram-Schmidt 
procedure) $\ket{\bar u_2}$ to $\ket{\hat u_1}$ and 
$\ket{\bar v_2}$ to $\ket{\hat v_1}$ resulting in 
new vectors $\ket{\hat u_2}$ and $\ket{\hat v_2}$ respectively. This 
procedure provides the QR-decomposition:
\begin{equation}
\label{eqQR}
(\ket{\bar u_1},\,\ket{\bar u_2})=(\ket{\hat u_1},\,\ket{\hat u_2})\,R_u
\end{equation}
where $R_u$ is the upper 
triangular $2\times 2$ matrix~:
\begin{equation}
\label{eqRu}
R_u=\left(\begin{array}{cc}
\|\ket{\bar u_1}\| & \langle \hat u_1\ket{\bar u_2} \\
0 &  \|\ket{\bar u_2}-\langle \hat u_1\ket{\bar u_2} \ket{\hat u_1}\|\\
\end{array}\right)\ .
\end{equation}
For $\ket{\bar v_{1,2}}$ and $\ket{\hat v_{1,2}}$ a similar relation 
holds using 
an upper triangular matrix $R_v$. Then the state $\ket{\psi(t+1)}$ 
after one iteration 
can be formally written as:
\begin{equation}
\label{eqstate1}
\ket{\psi(t+1)}=(\ket{\hat u_1},\,\ket{\hat u_2})\,R_u\,\hat \alpha(t)\,
R_v^T\left(\begin{array}{c}
\ket{\hat v_1}^T  \\
\ket{\hat v_1}^T  \\
\end{array}\right)
\end{equation}
where $\hat\alpha(t)$ is a $2\times 2$ diagonal matrix with entries 
$\alpha_1(t)$ and $\alpha_2(t)$ and where the notation 
$\ket{\hat v_{1.2}}^T$ indicates row vectors (due to the tensor product). 
This expression does not yet provide 
the Schmidt decomposition of $\ket{\psi(t+1)}$. For this it is necessary 
to compute the singular value decomposition of the $2\times 2$ matrix
\begin{equation}
\label{eqAmat}
A=R_u\,\hat \alpha(t)\,R_v^T=O_u\,\hat\alpha(t+1)\,O_v^T
\end{equation}
where $O_u$ and $O_v$ are unitary $2\times 2$ matrices and 
$\hat\alpha(t+1)$ is a diagonal matrix containing the new singular 
values $\alpha_1(t+1)$ and $\alpha_2(t+1)$. 
The new Schmidt vectors at time $(t+1)$ are then obtained by:
\begin{equation}
\label{eqnewschmidt}
(\ket{u_1(t+1)},\,\ket{u_2(t+1)})=(\ket{\hat u_1},\,\ket{\hat u_2})\,O_u
\end{equation}
and similarly for $\ket{v_{1,2}(t+1)}$ using $O_v$. 
This procedure shows that in absence of interaction 
the number of Schmidt components cannot increase and it provides 
also an efficient numerical method for the case $U=0$. During 
the iteration the renormalization due to the absorption can be done 
by renormalizing the obtained new singular values (such that 
$\alpha_1^2(t+1)+\alpha_2^2(t+1)=1$). 

Furthermore, for large times 
$\alpha_2$ decays exponentially while $\alpha_1\to 1$ 
(after renormalization) but fortunately $\alpha_2$ 
can be numerically computed in a very stable way, even if 
$\alpha_2\ll \alpha_1$, using 
the determinant of the matrix (\ref{eqAmat}) which is 
$|\det(A)|=|\det(R_u)\det(R_v)|\alpha_1(t)\,\alpha_2(t)
=\alpha_1(t+1)\,\alpha_2(t+1)$. 
First one computes $\alpha_1^2(t+1)$ as the leading eigenvalue 
of the matrix $A^\dagger A$ which is numerically stable and 
then $\alpha_2(t+1)$ is obtained from 
$\alpha_2(t+1)=|\det(A)|/\alpha_1(t+1)$.
This is also accurate for the case where the lower corner elements of 
the matrices $R_u$ and $R_v$ become very small which is possible 
due to the absorption process which has for long times $t\to\infty$ 
the tendency to produce 
vectors $\ket{\bar u_{1,2}}$ which are nearly parallel (and 
similarly for $\ket{\bar v_{1,2}}$).

\normalsize

\newpage
\section*{Appendix S: Supplementary Material for \\
``Loschmidt echo and Poincar\'e recurrences of
  entanglement''\\
by L.Ermann, K.M.Frahm and D.L.Shepelyansky}

\setcounter{equation}{0}
\renewcommand{\theequation}{S\arabic{equation}}
\setcounter{figure}{0}
\renewcommand\thefigure{S\arabic{figure}}
\renewcommand{\figurename}{SuppMat Figure}
\setcounter{page}{1} 

\label{apps}
         
Here we present additional Figures for the main part of the article.

Additional results related to the case of absorption of two particles 
at the chaos parameter value $K=7$ (discussed in Section 4) 
are presented in Fig.~S1.

In order to analyze the symmetry of the limit state (absorption of 
two particles at $K=7$) with respect to 
particle exchange, we decompose the state in an anti-symmetric component 
and a symmetric component by:
\begin{eqnarray}
\label{eqASym}
\psi_{\rm asym}(p_1,p_2)&=&\frac12\left(\psi(p_1,p_2)-\psi(p_2,p_1)\right)\\
\label{eqSym}
\psi_{\rm sym}(p_1,p_2)&=&\frac12\left(\psi(p_1,p_2)+\psi(p_2,p_1)\right)
\end{eqnarray}
and compute the norm of both components (after renormalization of the state 
$\ket{\psi}$ itself). It turns out that for all four interaction cases the 
symmetric component decays exponentially as 
can be seen in Fig.~S1 (for the case $U=2$, $U_r=1$; the 
other cases being similar) 
showing the time dependence of the norm $\|\ket{\psi_{\rm sym}(t)}\|$
and also of the difference $|\alpha_1-\alpha_2|$ of the 
first two singular values. The latter 
decays, apart from some oscillatory behavior, 
in the same way as the symmetric component showing the 
link between the anti-symmetry and the pairwise degeneracy of the 
singular values of the limit state for $t\to\infty$.

We mention that for $U=0$, the limit state (with only one Schmidt component) 
is symmetric with respect to particle exchange. We have also 
numerically verified that this symmetry persists for very small interactions 
values $U=\pm T/10$, $U_r=0,1$ where $T=\hbar=56/N$ is the parameter 
for the free rotation part of the kicked rotator (see discussion at the 
beginning of section \ref{sec4}). 
Apparently, for sufficiently strong interaction (with both signs) the 
leading mode of the non-unitary iteration operator is anti-symmetric with 
respect to particle exchange which is in contrast to the behavior of vanishing 
or very small interaction. 

\begin{figure}
\begin{center}
\includegraphics[width=0.8\textwidth]{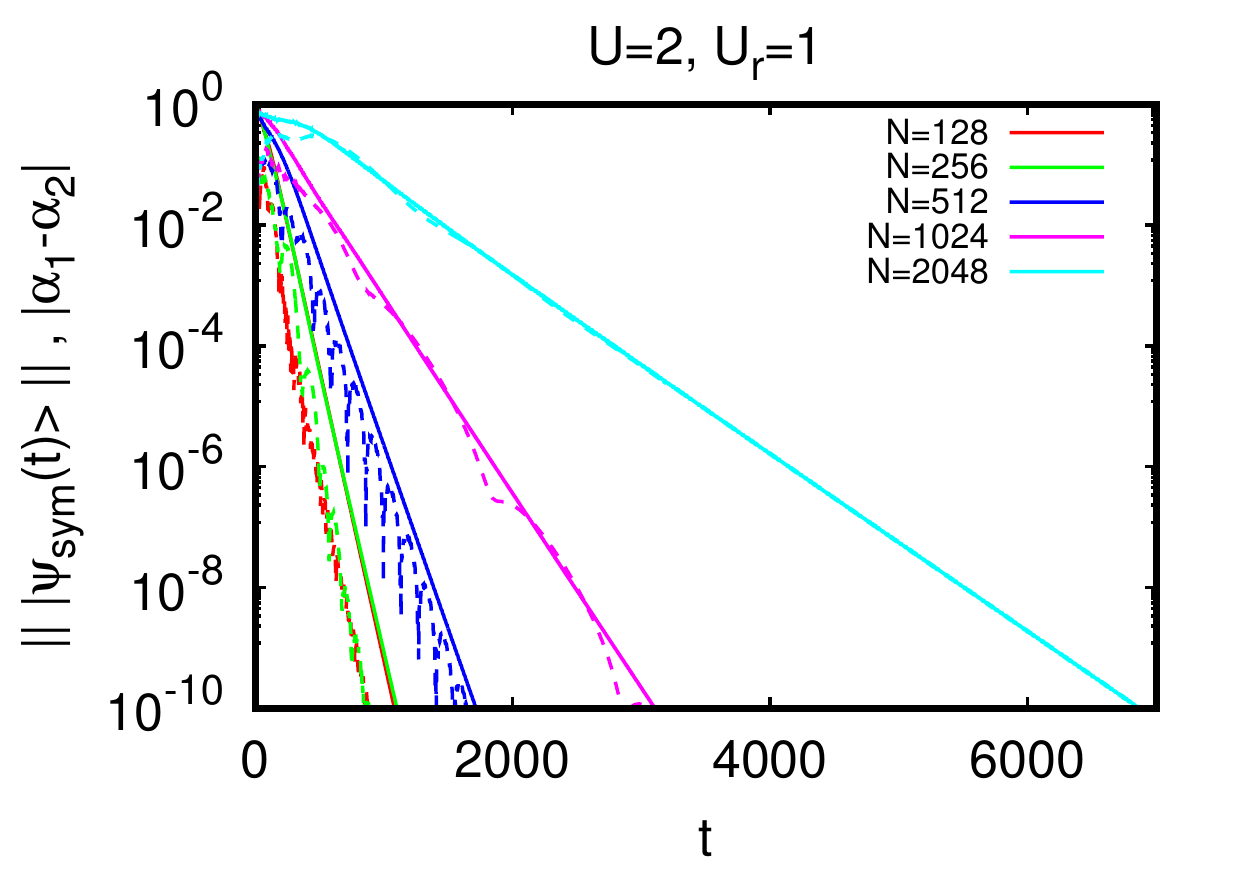}
\caption{
The full lines show the decay of the 
norm $\||\psi_{\rm sym}(t)\rangle\|$ of the particle 
symmetrized state $|\psi_{\rm sym}(t)\rangle$ obtained by 
$\psi_{\rm sym}(x_1,x_2)=(\psi(x_1,x_2)+\psi(x_2,x_1))/2$ 
(after renormalization of $|\psi(t)\rangle$)
for interaction values $U=2$, $U_r=1$ and system size $128\le N\le 2048$.
The dashed lines show for each value of $N$ the decay  of the difference 
$|\alpha_1-\alpha_2|$  where $\alpha_{1,2}$ are the first two singular 
values of the state $\ket{\psi(t)}$ becoming degenerate for $t\to\infty$.
}
\label{figS1}
\end{center}
\end{figure}

Additional results for the case of absorption of
only one particle at the chaos parameter value $K=2.5$
(discussed in Section 6) are presented in 
Figs.~S2, S3, S4.

The large values of $S_\infty$ (for $K=2.5$ and $N\ge 256$) 
imply that many singular values 
contribute to the state at final iteration time $t_{\rm max}=2^{18}$ 
as can be seen in Fig.~S2. In contrast to the cases with $K=7$,
there is (for $N\ge 256$) a considerable fraction 
of $\alpha_i\gtrsim 0.1$ values (roughly for $i\lesssim\sqrt{N/2}$) 
which contribute more uniformly to the two-particle state. Only 
for the case $N=128$ the first singular value $\alpha_1$ is 
somewhat dominating. 

\begin{figure}
\begin{center}
\includegraphics[width=0.8\textwidth]{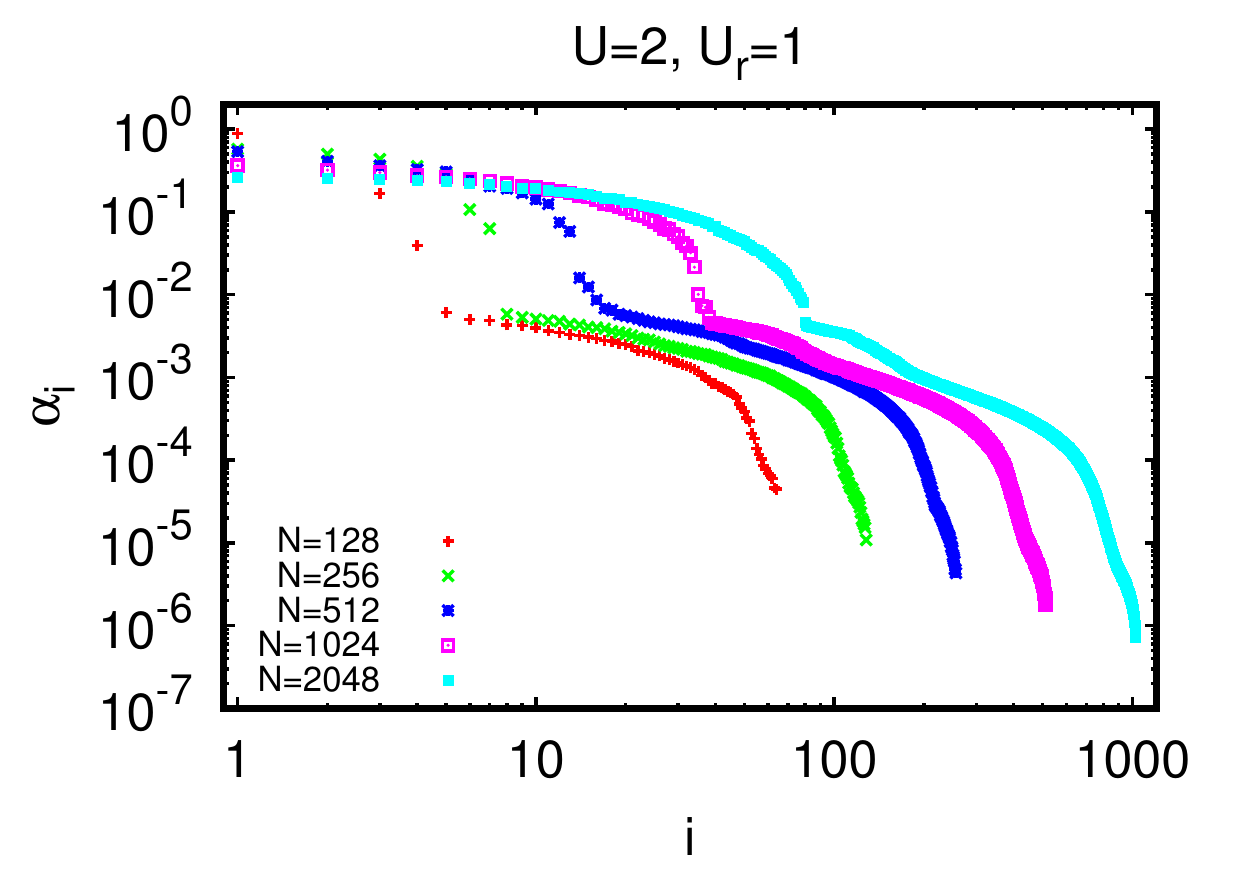}
\caption{
Singular values $\alpha_i$ (appearing in the  Schmidt decomposition of 
the state $|\psi(t=t_{\rm max})\rangle$ at the final 
iteration time $t_{\rm max}=2^{18}$) 
versus index $i$
for interaction values $U=2$, $U_r=1$, system size $128\le N\le 2048$, 
$K=2.5$  
and absorption only for the second particle. 
Both axis are shown on a logarithmic scale. The values of the top 
singular value are 
$\alpha_1=0.90055,\,0.57604,\,0.53490,\,0.36280,\,0.26346$ 
for $N=128,\,256,\,512,\,1024,\,2048$ respectively.
(The red data point for $N=128$ at $i=2$ with $\alpha_2=0.43810$ is 
hidden by the blue data point for $N=256$ with $\alpha_2=0.40901$.)
}
\label{figS2}
\end{center}
\end{figure}

\begin{figure}
\begin{center}
\includegraphics[width=0.8\textwidth]{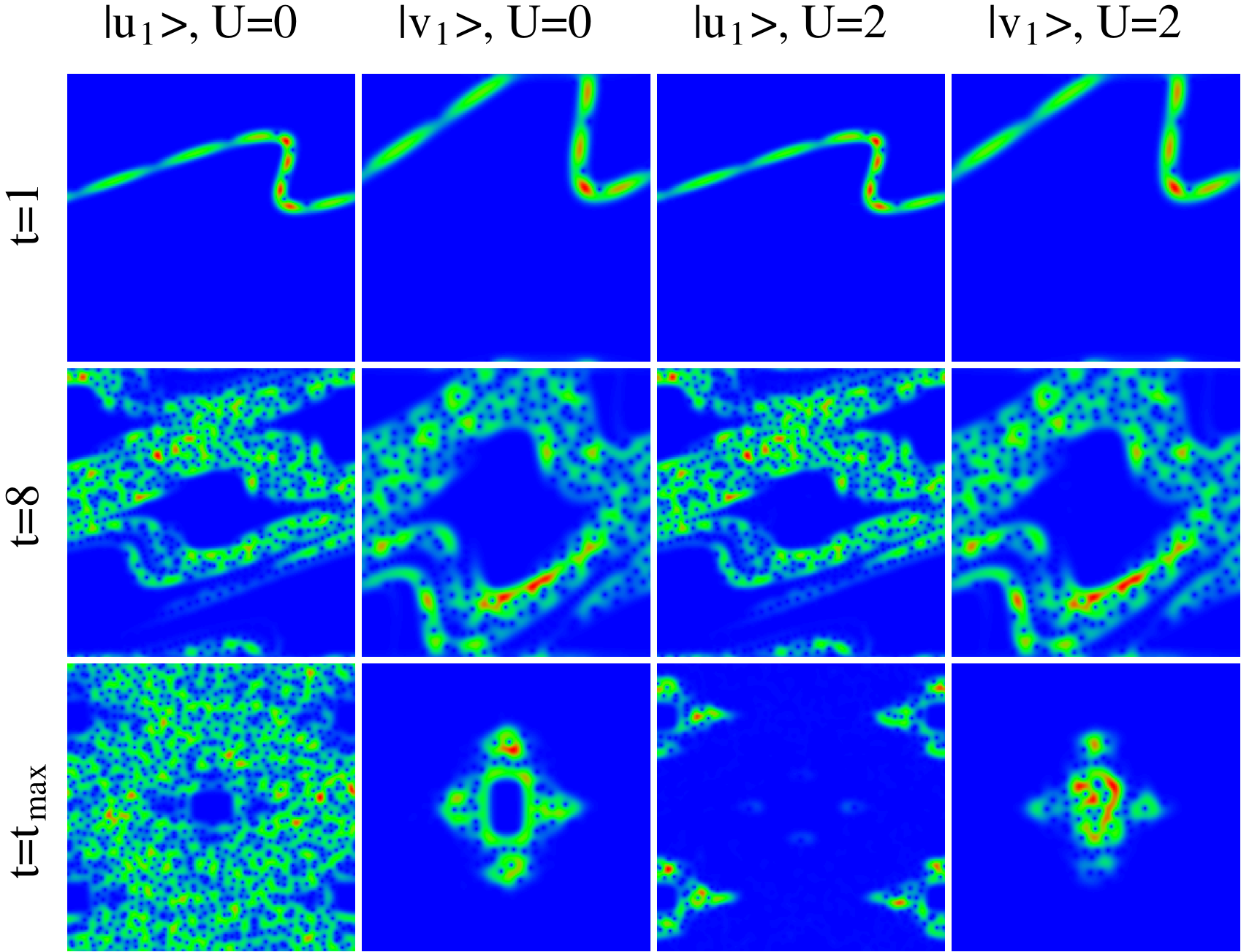}
\caption{
Husimi functions of the Schmidt states $|u_1\rangle$, $|v_1\rangle$, 
for $U=0$ or $U=2$ (with $U_r=1$), $N=1024$, $K=2.5$  
and absorption only for the second particle 
at three iteration times $t=1$, $t=8$ and 
$t=t_{\rm max}=2^{18}$ for both $U=0$ and $U=2$. 
The horizontal axis corresponds to the phase  $\theta\in[0,2\pi]$
(different representation from former Husimi figures)  and 
the vertical axis to the momentum $p_{\rm cl}\in[-2p_{\rm max},2p_{\rm max}]$ 
(for $|u_1\rangle$) or $p_{\rm cl}\in[-p_{\rm max},p_{\rm max}]$ 
(for $|v_1\rangle$) where $p_{\rm max}=5=(2\pi)\cdot 0.7958$ 
is the momentum absorption border in classical units (for the second 
particle) and $2p_{\rm max}$ is the maximal 
classical momentum value for the full phase
space in absence of absorption (for the first particle). 
The colors red/green/blue correspond to 
maximal/medium/minimal values of the Husimi function. 
}
\label{figS3}
\end{center}
\end{figure}

As for the cases with $K=7$, we have also computed for $K=2.5$ the first 
Schmidt vectors $\ket{u_{1,2}}$, $\ket{v_{1,2}}$ and the associated 
Husimi functions \cite{husimi,qcfrahm,eprchaos}. 
The vectors $\ket{u_1(t)}$, $\ket{v_1(t)}$ for $U=0$ and 
$U=2$, $U_r=1$ (all for $N=1024$) 
are shown in Fig.~S3 for three time values 
being $t=1$, $t=8$ and $t=t_{\rm max}=2^{18}$ 
for both $U=0$ and $U=2$, $U_r=1$. 
Again for $t=1$, these states occupy the same manifold but now 
there is no visible difference between $U=0$ and $U=2$. 
Note that the available phase space for $\ket{v_1(t)}$ is reduced by a factor 
of 2 concerning the maximal $p$ value as compared to $\ket{u_1(t)}$ 
and that in Fig.~S3 the horizontal axis corresponds 
to $\theta\in[0,2\pi]$ in order to be coherent with the Husimi functions 
shown in \cite{maspero2} (instead of $\theta\in[-\pi,\pi]$ in the 
Husimi figures for $K=7$). 

Again for $t=8$ the phase space structure is quite complicated but rather 
similar between the interacting and non-interacting cases for both 
states $\ket{u_1}$ and $\ket{v_1}$ respectively. 

At long times the state $\ket{u_1}$ of $U=0$ seems to be ergodic 
in the chaotic region of the phase space together with visible 
strong islands while the state $\ket{v_1}$ of $U=0$ is concentrated around 
the stable islands but without penetrating the main center island 
(the other islands are quite small on the scale of the resolution 
of the Husimi function for $N=1024$). 
For $U=2$ the state $\ket{u_1}$ is localized at four structures close to the 
corners indicating a strong repulsion of the first particle from the
second particle due to the interaction. 
For $U=2$ the state $\ket{v_1}$ is localized around 
the main stable island but now it penetrates inside the island quite strongly. 
It seems that the interaction reduces the penetration time which may be linked 
to the slow but well visible exponential decay of $S(t)$ to $S_\infty$.

\begin{figure}
\begin{center}
\includegraphics[width=0.8\textwidth]{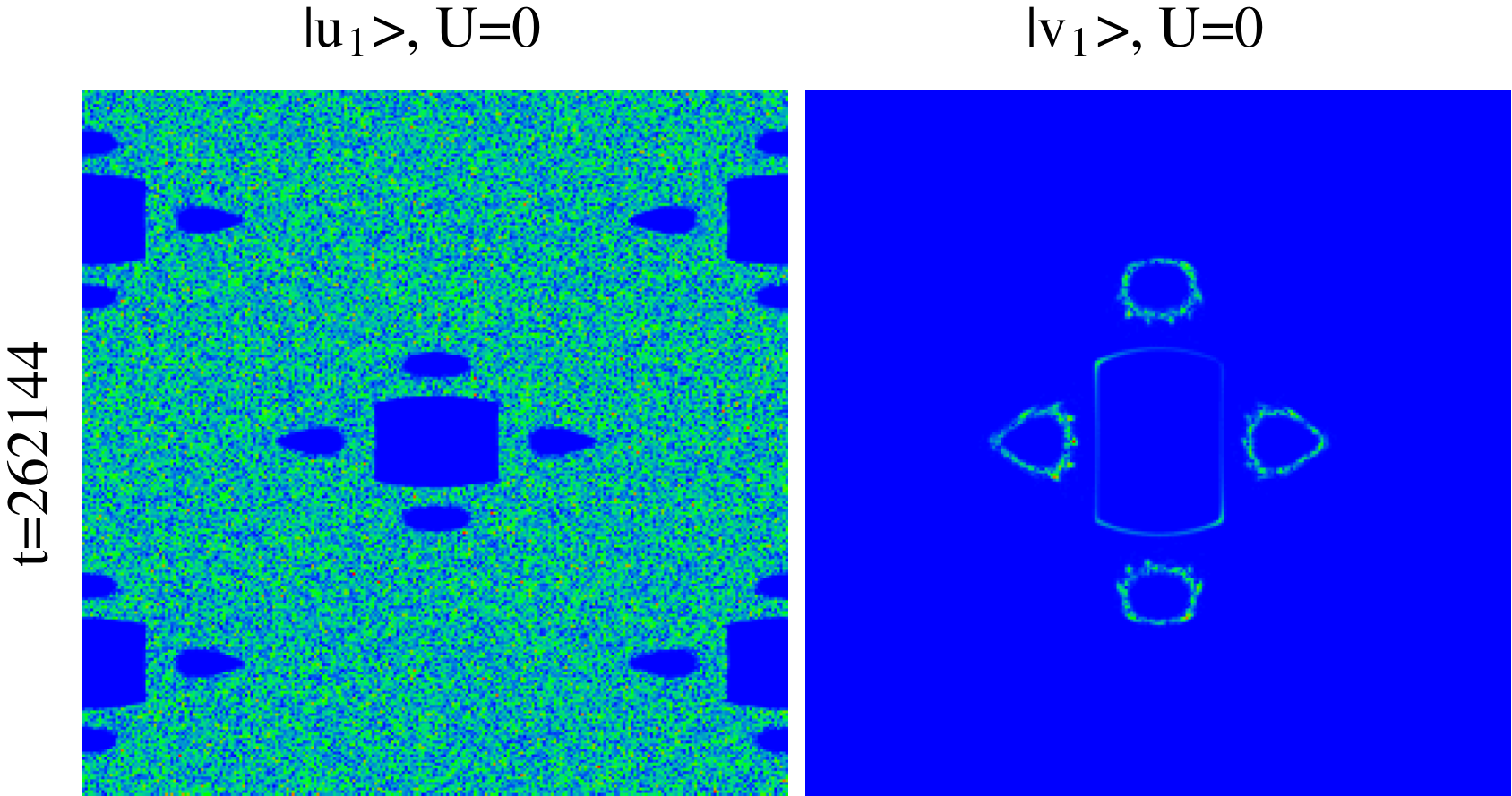}
\caption{
Husimi functions of the Schmidt states $|u_1\rangle$, $|v_1\rangle$, 
for $U=0$, $N=65536$, $K=2.5$  
and absorption only for the second particle 
at final iteration time $t=2^{18}$. The signification 
of both axes and the color codes is the same as in Fig.~S3.
}
\label{figS4}
\end{center}
\end{figure}

In Fig.~S3, we also show for the asymmetric absorption case 
with $K=2.5$ 
the Husimi functions of $\ket{u_1}$ and $\ket{v_1}$
for $U=0$ and a larger system size $N=65536$ at final iteration time 
$t=2^{18}$.
Similarly to the case $N=1024$ the state $\ket{u_1}$ is ergodic in the 
chaotic region of the phase space with very distinctly visible stable 
islands and $\ket{v_1}$ is strongly localized around the center island 
(with modest density) and around the four secondary islands 
(with stronger density) but without penetrating these islands. 
The figure of $\ket{v_1}$ is similar to the Husimi figures of Fig.~4 
of \cite{maspero2} (which correspond to $N=13222$ in our notation for 
$N$).

\end{document}